\documentclass[dvips,11pt,a4paper]{article} 
\usepackage{amsmath}
\usepackage{graphicx}
\usepackage[bookmarks=true,colorlinks=true,linkcolor=black,citecolor=blue]{hyperref}
\usepackage{bm}
\usepackage{authblk}
\reversemarginpar
\newcommand\TT{\rule{0pt}{2.5ex}}        
\newcommand\BB{\rule[-1.0ex]{0pt}{0pt}}  
\DeclareMathOperator{\arctanh}{arctanh}

\newcommand{\be}{\begin{equation}}
\newcommand{\en}{\end{equation}}
\newcommand{\bea}{\begin{eqnarray}}
\newcommand{\ena}{\end{eqnarray}}
\newcommand{\bp}{\begin{pmatrix}}
\newcommand{\ep}{\end{pmatrix}}
\newcommand{\lbl}[1]{\label{eq:#1}}
\newcommand{\lbltab}[1]{\label{tab:#1}}
\newcommand{\lblfig}[1]{\label{fig:#1}}
\newcommand{\lblsec}[1]{\label{sec:#1}}
\newcommand{\rf}[1]{(\ref{eq:#1})}
\newcommand{\Table}[1]{\ref{tab:#1}}
\newcommand{\fig}[1]{\ref{fig:#1}}
\newcommand{\sect}[1]{\ref{sec:#1}}
\newcommand{\braque}[1]{{\langle #1 \rangle}}
\newcommand{\ket}[1]{{\vert #1 \rangle}}

\newcommand{\bc}{\begin{center}}
\newcommand{\ec}{\end{center}}
\newcommand{\bt}{\begin{tabular}}
\newcommand{\et}{\end{tabular}}
\newcommand{\ba}{\begin{array}}
\newcommand{\ea}{\end{array}}
\newcommand{\MeV}{\hbox{MeV}}
\newcommand{\im}{{\rm Im\,}}
\newcommand{\re}{{\rm Re\,}}
\newcommand{\disc}{{\rm disc}}
\newcommand{\Kbar}{\bar{K}}
\newcommand{\mpi }{m_\pi}
\newcommand{\fpi }{F_\pi}
\newcommand{\meta }{m_\eta}
\newcommand{\mpid}{m_\pi^2}

\newcommand{\fpid}{F_\pi^2}

\newcommand{\mkd}{m^2_K}
\newcommand{\mkpd}{m^2_{K^+}}

\newcommand{\Kp}{{K^+}} 
\newcommand{\Km}{{K^-}}
\newcommand{\Kz}{{K^0}} 
\newcommand{\Kzb}{{\bar{K}^0}}
\newcommand{\piz}{{\pi^0}} 
\newcommand{\pip}{{\pi^+}}
\newcommand{\pim}{{\pi^-}}
\newcommand{\metad}{m_\eta^2}

\newcommand{\mvd}{m_V^2}

\newcommand{\qd}{q^2}

\newcommand{\gapprox}{%
\mathrel{%
\setbox0=\hbox{$>$}\raise0.6ex\copy0\kern-\wd0\lower0.65ex\hbox{$\sim$}}}
\newcommand{\lapprox}{%
\mathrel{%
\setbox0=\hbox{$<$}\raise0.6ex\copy0\kern-\wd0\lower0.65ex\hbox{$\sim$}}}

\date{\today}
\title{Revisiting $\gamma^*\to\gamma\pi^0\eta$ near the  $\phi(1020)$
using analyticity and the left-cut structure}

\author[1]{B. Moussallam}

\affil[1]{\small Laboratoire Ir\`ene Joliot-Curie
  (CNRS/IN2P3, UMR9012), P\^ole Th\'eorie, Universit\'e  
  Paris-Saclay, 91406 Orsay, France}

\begin{document}

\maketitle

\begin{abstract}
Amplitudes of the form $\gamma^*(\qd)\to \gamma P_1P_2$ appear as sub-processes
in the computation of the muon $g-2$. We test a proposed theoretical modelling
against very precise experimental measurements by the KLOE collaboration at
$\qd=m^2_\phi$. Starting from an exact parameter-free dispersive
representation for the $S$-wave satisfying QCD asymptotic constraints and
Low's soft photon theorem we derive, in an effective theory spirit, a
two-channel Omn\`es integral representation which involves two subtraction
parameters. The discontinuities along the left-hand cuts which, for timelike
virtualities, extend both on the real axis and into the complex plane are
saturated by the contributions from the light vector mesons. In the case of
$P_1P_2=\pi\eta$, we show that a very good fit of the KLOE data can be
achieved with two real parameters, using a $T$-matrix previously determined
from $\gamma\gamma$ scattering data. This indicates a good compatibility
between the two data sets and confirms the validity of the $T$-matrix. The
resulting amplitude is also found to be compatible with the chiral soft pion
theorem.  Applications to the $I=1$ scalar form factors and to the $a_0(980)$
resonance complex pole are presented.
\end{abstract}

\tableofcontents

\section{Introduction}
The discrepancy between the experimental determinations of the muon $g-2$ and
its calculation within the Standard Model (e.g.~\cite{ Jegerlehner:2009ry})
has recently been confirmed by a new measurement~\cite{Abi:2021gix}.  Since
the soft hadronic contribution to the $g-2$ has the largest error in the
calculation it seems necessary to further investigate the theoretical
descriptions of amplitudes involving light mesons together with real or
virtual photons and to also further improve our knowledge of the interactions
of these mesons among themselves and the properties of the associated light
resonances.

We reconsider here the description of amplitudes of the form $\gamma^*(\qd)
\to \gamma P_1 P_2$ involving one real and one virtual photon. In the context
of the $g-2$, the channel $P_1 P_2=\pip\pim$ is the most relevant one. It was
considered in refs.~\cite{Colangelo:2017fiz,Hoferichter:2019nlq}, also in the
case of two virtual photons, using a method which applies when the energy of
the $P_1P_2$ system is much smaller than 1 GeV. Recently, a
generalisation was performed~\cite{Danilkin:2021icn}, from which the
contribution of the $f_0(980)$ resonance to the muon $g-2$ was estimated.
We develop here a similar approach, in the case of one virtual photon, which
can be applied in an energy region of the $P_1P_2$ system
slightly larger than 1 GeV.
We focus on the case where
$P_1 P_2 =\piz\eta$, $(K\Kbar)_{I=1}$ and will test the method against the
very precise measurements performed by the KLOE collaboration close to the
$\phi(1020)$ meson peak~\cite{Ambrosino:2009py} (earlier results can be found
in~\cite{Aloisio:2002bsa,Achasov:2000ku,Akhmetshin:1999di,Achasov:1998cc}).
The amplitudes $\phi\to \gamma\piz\eta$, $\phi\to \gamma\piz\piz$ have large
resonant contributions induced by the scalar mesons $a_0(980)$, $f_0(980)$
respectively. Initial interest in measuring such amplitudes was stimulated by
the claim~\cite{Achasov:1987ts} that this would allow to clearly discriminate
between models of these presumably exotic mesons (the
$q^2\bar{q}^2$~\cite{Jaffe:1976ig} versus the $K\Kbar$ molecule
model~\cite{Weinstein:1982gc,Weinstein:1983gd}) but some disagreements with
this claim have been expressed (e.g.~\cite{Kalashnikova:2004ta}). We refer
to~\cite{Klempt:2007cp} for a review on this topic.

Our objective here is rather to probe both the quality of the description of
the radiative amplitude which can be achieved and the properties of the
$\pi\eta$ $S$-wave $T$-matrix globally in the low/medium energy range. While
the existence of a sharp $I=1$ scalar resonance coupling to both $\pi\eta$ and
$K\Kbar$ channels was established long ago~\cite{Astier:1967zz,Ammar:1969vy}
the detailed behaviour of the phase shifts is still controversial. The pattern
which was established in lattice QCD simulations~\cite{Dudek:2016cru} at
$\mpi=391$ MeV may or may not extend down to the physical pion mass value,
depending on the extrapolation model~\cite{Guo:2016zep}. A very similar
phase shift pattern was found to emerge in the meson-meson scattering model
developed in ref.~\cite{Danilkin:2011fz} which was applied to $\gamma\gamma\to
\pi\eta$ scattering in ref.~\cite{Danilkin:2017lyn} and found to describe
reasonably well the data by the Belle
collaboration~\cite{Uehara:2009cf}. However, the mass and width properties of
the $a_0(980)$ resonance in this model are not compatible with the PDG
values. Moreover, a determination of the $\pi\eta-K\Kbar$ scattering matrix
performed in ref.~\cite{Lu:2020qeo} using the same set of $\gamma\gamma$ data,
together with data on $\gamma\gamma\to K_S K_S$ from~\cite{Uehara:2013mbo} and
$\gamma\gamma\to \Kp\Km$ from~\cite{Albrecht:1989re} obtained different
results for the phase shifts and for the properties of the $a_0$ resonance.

Our approach is based on the general analyticity structure of the
$\gamma^*\to \gamma \pi\eta, K\Kbar$ partial waves and the use of the
Omn\`es method~\cite{Omnes:1958hv} (extended to several
channels~\cite{Bjorken:1960zz}) which ensures, by construction, that
two-channel unitarity is exactly satisfied. In some previous
work~\cite{Achasov:1987ts, Achasov:2002ir,Isidori:2006we} the emphasis
has been on a quick determination of the $a_0(980)$ parameters, but
the amplitude models eventually violate unitarity, which may introduce
a bias in this determination. The amplitude models used in
refs.~\cite{Marco:1999df,Palomar:2003rb} based on a unitarisation of
the leading order chiral expansion (u$\chi$PT) do fulfil unitarity
(the relation to our formalism is worked out in
appendix~\sect{kaonloopderiv}) but the underlying $T$-matrix is
possibly somewhat unrealistic (having e.g. no left-hand cuts).

We develop an Omn\`es-type representation for $\gamma^*\to \gamma\pi\eta,
\gamma K\Kbar$ $S$-waves which involves explicit integrations over the
left-cut. When the virtuality $\qd$ is negative or vanishing, this left-cut
has a  simple structure, lying on the negative real axis, but this changes
for timelike virtualities.  In that case, the left-cut has several components
extending in the complex plane~\cite{Kennedy:1961}. The integrations have to
be done carefully and we explain this in detail. One component of this
left-cut eventually overlaps with part of the unitarity cut and turns around
the threshold point. In addition the Born amplitude has a pole singularity
which also lies on top of the unitarity cut.  These singularities induce a
deviation of the phase of the $\gamma^*\to \gamma\pi\eta$ partial wave from
the the $\pi\eta$ elastic phase shift such that Watson's theorem does not
apply~\cite{Creutz:1969hv}.  A further consequence concerns the Adler zero
which, in the partial wave, is moved to an unphysical Riemann sheet.

This coupled-channel dispersive representation involves two real subtraction
parameters. The presence of subtraction parameters is a necessary and
unavoidable consequence of the effective-theory nature of dispersion relations
and should not be omitted, they absorb required corrections to the higher
energy parts of the integrations. The minimal number of subtractions is
determined from the asymptotic bounds on the partial waves and the dynamical
constraints which must be satisfied, like the exact soft photon zero in the
present case~\cite{Low:1958sn}. Because of the very small number of
undetermined parameters which are involved, the amplitudes $\gamma^*\to
\gamma\pi\eta$, similarly to $\gamma\gamma\to \pi\eta$, are very good probes
of the final-state interaction $T$-matrix. We
will show that a rather good fit of the KLOE data~\cite{Ambrosino:2009py} on
$\phi\to\gamma\pi\eta$ can be achieved with two fit parameters, using a
$T$-matrix previously determined from $\gamma\gamma$
data~\cite{Lu:2020qeo}. We will also study how combining the two sets of data
can help improve the determination of the $T$-matrix and the properties of the
$a_0(980)$ resonance.

The plan of the paper is as follows. After recalling some general
properties related to gauge invariance we discuss the origin of the
cuts in the partial waves and the approximations to be used for
evaluating the discontinuities, based on vector meson exchanges in
crossed channels.  Starting form a general, exact, dispersive
representation for the partial waves of interest we derive a
coupled-channel Omn\`es representation valid in the considered energy
region. This representation involves two subtraction parameters and a
number of integrals over both the unitarity cut and the left cuts. We
then discuss the evaluation of the vector meson coupling constants
which are needed (including their relative signs) using experimental
inputs together with flavour, chiral and asymptotic constraints. After
explaining how to accurately compute all the integrals over the
complex contours we perform a detailed comparison with the
experimental data and discuss some consequences.

\section{General formalism}
\subsection{Tensor and helicity amplitudes}
We first recall that amplitudes involving two real or virtual photons
and two pseudo-scalar mesons $\gamma^*(q_2)\to \gamma^*(q_1) P_1(p_1)
P_2(p_2)$ can be derived from a correlation function involving two
electromagnetic currents
\be
W_{\mu\nu}(p_1,p_2;q_1,q_2)=
i\int d^4x d^4y \exp(-iq_1x+iq_2y) \braque{P_1P_2\vert
T[j^{em}_\mu(x) j^{em}_\nu(y)] \vert 0}
\en
Current conservation $\partial^\alpha j^{em}_\alpha=0$ implies that
the tensor $W_{\mu\nu}$ must satisfy the two Ward identities
\be\lbl{Wardq1q2}
q_1^\mu W_{\mu\nu} = q_2^\nu  W_{\mu\nu} =0\ .
\en 
One can then expand $W_{\mu\nu}$ over a basis of five independent
tensors (e.g.~\cite{Bardeen:1969aw}) satisfying~\rf{Wardq1q2} formed
with the three independent momenta $q_1$, $q_2$ and $\Delta=p_1-p_2$,
\be\lbl{5tensors}
\ba{l}
T_{1\mu\nu}= q_1\cdot q_2\, g_{\mu\nu} -q_{1\nu} q_{2\mu}\\[0.2cm]
T_{2\mu\nu}= 4\Delta_\mu     (q_1\cdot q_2\,
\Delta_\nu-q_2\cdot\Delta\, q_{1\nu}) 
            -4q_1\cdot\Delta (q_{2\mu}\Delta_\nu
            -q_2\cdot\Delta\, g_{\mu\nu})\\[0.2cm]
T_{3\mu\nu}= 2\Delta_\mu    (q_1\cdot q_2\,  q_{2\nu}-q_2^2\, q_{1\nu})
            -2q_1\cdot\Delta(q_{2\mu} q_{2\nu} - q_2^2\, g_{\mu\nu})\\[0.2cm]
T_{4\mu\nu}= q_{1\mu}(q_{1\nu} q_2^2-q_{2\nu} q_1\cdot q_2)
            +q_1^2(q_{2\mu}q_{2\nu}-q_2^2\,g_{\mu\nu})\\[0.2cm]
T_{5\mu\nu}= q_{1\mu}(q_{1\nu} q_2.\Delta-\Delta_\nu q_1\cdot q_2)
             +q_1^2(q_{2\mu}\Delta_\nu-q_2\cdot\Delta\,g_{\mu\nu})
\ea
\en
We consider here more specifically the situation where one photon is
real and the other is virtual,
\be
q_1^2=0,\quad q_2^2\equiv q^2 \ne 0\  .
\en
In this case, the tensors $T_{4\mu\nu}$, $T_{5\mu\nu}$ are not
physically relevant.

The dependence as a function of the virtuality
$\qd$ is expected to display large Breit-Wigner peaks associated with
the light vector resonances $\rho$,$\omega$,$\phi$, such that
\be\lbl{q2nearphi}
W^{\mu\nu}(p_1,p_2;q_1,q_2)\simeq 
\sum_{V'=\rho,\omega,\phi,\cdots}
\frac{m_{V'}^2\,f_{V'}}{\qd - m_{V'}^2 +im_{V'} \Gamma_{V'}}
{\cal T}_{V'}^{\mu\nu}(p_1,p_2;q_1,q_2)
\en
where ${\cal T}_{V'}^{\mu\nu}$ describes the amplitude for the vector meson
$V'$ to decay into $\gamma P_1 P_2$. We will consider here the situation where
$\qd$ is close to the peak of the $\phi(1020)$ resonance i.e. $q^2\simeq
m^2_\phi$, and focus on the construction of the amplitude ${\cal
  T}_\phi^{\mu\nu}$, assuming that the background contributions from the the
other vector resonances can be neglected or have been subtracted.  In the
sequel, the index $\phi$ will be dropped. This amplitude can be
expanded in terms of the independent tensors as 
\be
{\cal T}^{\mu\nu} = A(s,t,u) T_1^{\mu\nu}+ B(s,t,u) T_2^{\mu\nu}
+C(s,t,u) T_3^{\mu\nu}
\en
where the functions $A$, $B$, $C$ are Lorentz scalars depending 
on the external masses and on the Mandelstam invariants $s$, $t$, $u$, 
\be
s=(p_1+p_2)^2,\quad
t=(p_1+q_1)^2,\quad
u=(p_2+q_1)^2
\en
which satisfy $s+t+u= \qd +m^2_1 +m^2_2$.  We can derive the helicity
amplitudes by contracting the tensorial amplitude with the
polarisation vectors of the photon and the $\phi$ meson
\be
e^{i(\lambda_2-\lambda_1)\phi} {\cal T}_{\lambda_1\lambda_2}(s,\theta)=
-e_\gamma^{*\mu}(q_1,\lambda_1) e_\phi^\nu(q_2,\lambda_2)
{\cal T}_{\mu\nu}\ .
\en
(with a conventional minus sign). The
scattering angle $\theta$ is defined as the angle between $\vec{p}_1$
and $\vec{q_1}$ in the centre-of-mass system of the two pseudoscalar
mesons (see fig.~\fig{m1m2cms}). The Mandelstam variables $t$, $u$
are expressed as follows as a function of $\theta$
\be
\ba{l}
t= m_1^2 +\dfrac{\qd-s}{2s} \left(s +m_1^2-m_2^2 
 -\lambda_{12}(s) \cos\theta\right) \\[0.3cm]
u=m_2^2 +\dfrac{\qd-s}{2s} \left(s +m_2^2-m_1^2 
 +\lambda_{12}(s) \cos\theta\right) 
\ea\en
where
\be
\lambda_{12}(s)=\sqrt{s^2-2s\Sigma_{12}+ \Delta_{12}^2}
\en
with
\be
\Sigma_{12}=m_1^2+m_2^2,\quad
\Delta_{12}=m_1^2-m_2^2\ .
\en
The three independent helicity amplitudes, finally, can be written 
in terms of the three functions $A$, $B$, $C$ as
\be\lbl{helicABC}
\ba{ll}
{\cal T}_{++}= & (\qd-s)\Big[\dfrac{1}{2}A(s,t,u)+\Big(2\Sigma_{12}-s
+\dfrac{\qd}{s^2}\big(\lambda_{12}(s)\cos\theta-\Delta_{12}\big)^2\Big)
\,B(s,t,u)\\[0.4cm]
\  & -\dfrac{\qd}{s}\big(\lambda_{12}(s)\cos\theta-\Delta_{12}\big)
\,C(s,t,u)\Big]\\[0.4cm]
{\cal T}_{+-}= & (\qd-s)\dfrac{\lambda^2_{12}(s)}{s}\sin^2\theta
\,B(s,t,u)\\[0.4cm]
{\cal T}_{+0}= & (\qd-s)\sqrt{\dfrac{\qd}{2 s}}\,{\lambda_{12}(s)}
\,{\sin\theta}\Big[
\dfrac{2}{s}\big(\lambda_{12}(s)\cos\theta-\Delta_{12}\big)\,B(s,t,u)
-C(s,t,u)\Big]\ .
\ea\en
\begin{figure}
\centering
\includegraphics[width=0.60\linewidth]{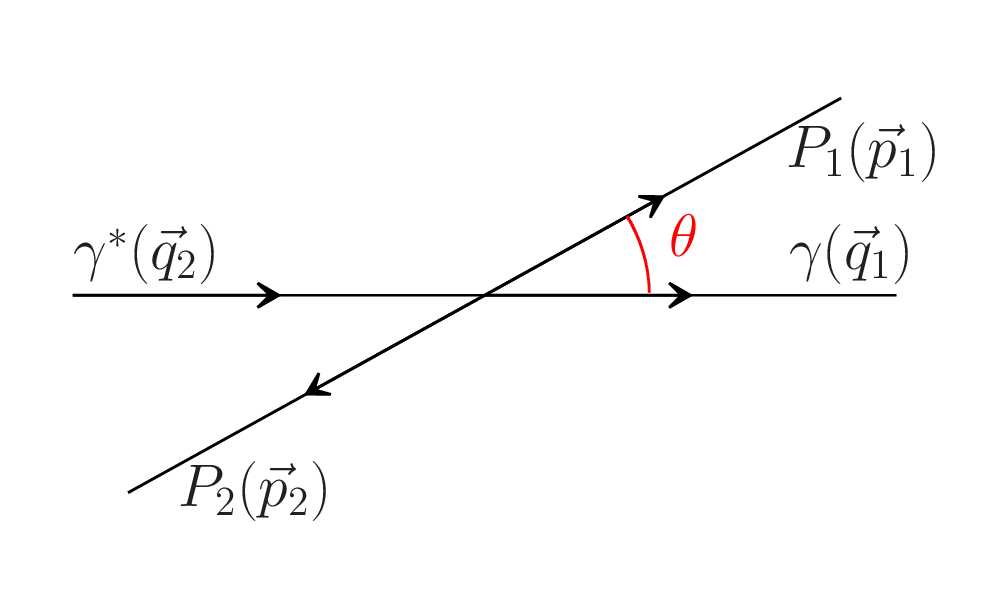}
\caption{\small  Centre-of-mass system for the $\gamma^*(q_2) \to \gamma(q_1)
  P_1(p_1) P_2(p_2)$ amplitude.}
\lblfig{m1m2cms}
\end{figure}
\subsection{Partial waves and their singularities}\lblsec{pwsingular}
We will designate the helicity amplitudes of interest as
\be
\ba{ll}
\phi \to \gamma\pi\eta :  & L_{\lambda\lambda'} \\[0.2cm] 
\phi \to \gamma\Kp\Km :   & K^c_{\lambda\lambda'} \\[0.2cm] 
\phi \to \gamma\Kz\Kzb :  & K^n_{\lambda\lambda'} 
\ea\en
The isospin $I=0,1$ $K\Kbar$ amplitudes are related to the $\Kp\Km$, 
$\Kz\Kzb$ ones by
\be
\ba{l}
K^0_{\lambda\lambda'}=-\dfrac{1}{\sqrt2}( K^c_{\lambda\lambda'}
                                  + K^n_{\lambda\lambda'})\\[0.3cm] 
K^1_{\lambda\lambda'}=-\dfrac{1}{\sqrt2}( K^c_{\lambda\lambda'}
                                  - K^n_{\lambda\lambda'})
\ea\en
and we consider here the coupled-channel system of the two $S$-waves
with $I=1$ 
\be
\ba{l}
l_{0++}(s)   =\dfrac{1}{2}{\displaystyle \int_{-1}^1} L_{++}(s,z)\,dz\\[0.3cm]
k^1_{0++}(s) =\dfrac{1}{2} {\displaystyle \int_{-1}^1} K^1_{++}(s,z)\,dz
\ea\en
The angular integration can be expressed in terms of the
$t$-variable, e.g. for the $\pi\eta$ amplitude
\be\lbl{tintegral}
l_{0++}(s)   =\dfrac{s}{(\qd-s)\lambda_{\pi\eta}(s)}
\int_{t^-(s)}^{t^+(s)} L_{++}(s,t)\,dt
\en
with
\be
t_\pm(s)=\mpid+ \frac{\qd-s}{2s}(s+\mpid-\metad\pm \lambda_{12}(s))\ .
\en
Let us first examine the singularities of the partial wave when
$s\to0$ and $s\to (\meta\pm \mpi)^2$:
\begin{itemize}
\item[{\bf 1)}] {\bf Singularity when $s\to0$:} When $s\to0$, the lower
  integration boundary in eq.~\rf{tintegral} goes to $\infty$ since one has
\be
\left.t^-(s)\right\vert_{s\to0}= \frac{\qd (\mpid-\metad)}{s}+ 
\metad\Big(1+\frac{\qd}{\metad-\mpid}\Big) +O(s)
\en
which leads to a singular behaviour of $l_{0++}(s)$
(see~\cite{Jakob:1969hn}).  The behaviour of the amplitude
$L_{++}(s,t)$ in the regime when $s\to 0$ and $t\to \infty$ is driven
by the $a_2$ Regge trajectory which gives,
\be
\left.L_{++}(s,t)\right\vert_{s\to0,t\to\infty}\sim \beta(s)\left(
\frac{t}{s_0}\right)^{\alpha_0+ \alpha' s}
\en
with $\alpha_0\simeq0.5$, $\alpha'\simeq 0.9$ $\hbox{GeV}^{-1}$. Using
this in eq.~\rf{tintegral} one obtains for the $S$-wave near $s=0$, 
\be
\left.l_{0++}(s)\right\vert_{s\to0}\sim \left( 
\frac{\qd (\metad-\mpid)}{s_0\, s}\right)^{\alpha_0+\alpha' s}\ .
\en
\item[{\bf 2)}]{\bf Singularities when $s\to (\meta\pm \mpi)^2$}:
When performing the $t$-integration~\rf{tintegral} with a fixed value
of the energy $s$ one must pay attention to the cuts of the function
$L_{++}(s,t)$ in the $t$ variable. In the $t$-channel, which we can
write $\phi\eta\to \gamma\pi$ the lightest unitarity contributions are
from isoscalar $3\pi$ states which generates a cut in the $t$ variable
lying on the real axis from $t=t_0=9\mpid$ to $t=\infty$. Similarly,
in the $u$ channel $\phi\pi\to \gamma\eta$ the lightest unitarity
contributions are from isovector $\pi\pi$ states. This generates a cut
in the $t$ variable extending from $-\infty$ to
$t(u_0,s)=\qd+\metad+\mpid-u_0-s$ with $u_0=4\mpid$. This cut can be
shifted away from the real axis by appending an infinitesimal imaginary
part to $\qd$
\be\lbl{q2+ieps}
\qd \to \qd+i\epsilon\ .
\en
When $s$ is in the physical region, $\lambda_{\pi\eta}(s)$ is real and one
easily verifies that
\be
0< \im[t_\pm(s)]< \epsilon\quad (s > (\meta+\mpi)^2) .
\en
The integration path in eq.~\rf{tintegral} therefore lies in between the two
cuts without touching them for finite values of $s$.
The situation changes when $s$ is in the unphysical region
$(\meta-\mpi)^2 < s < (\meta+\mpi)^2$ where $\lambda_{\pi\eta}(s)$ is
imaginary. The path of integration  must then be
distorted to turn around the cut as illustrated in fig.~\fig{zintpath}. As a
consequence, the partial wave $l_{0++}(s)$ diverges when $s$ approaches the
threshold from below since the integral in eq.~\rf{tintegral} remains finite
while $\lambda_{\pi\eta}(s)$ in the denominator vanishes. A similar divergence
occurs when $s$ approaches the pseudo-threshold $(m_\eta-m_\pi)^2$ from
above. The discontinuities of $l_{0++}(s)$ when $s$ moves across the points
$(m_\eta\pm m_\pi)^2$ reflect the fact that $s$ crosses the cut of the
partial-wave amplitude at these points. This cut will be described in more
detail below. 
\end{itemize}

The singularities of the partial waves when $s$ crosses the points
$(m_\eta\pm m_\pi)^2$ affects the position of the Adler zero.
The existence
of an Adler zero in the chiral limit when the pion is soft, i.e. $p_1=0$,
implies that the physical helicity amplitude $L_{++}(s,t)$ should have an
Adler zero at $t=\mpid$, $s=s_A\simeq\metad$. The Adler zero is also present
in the $J=0$ partial wave for small enough $q^2$
virtualities\footnote{The condition for $l_{0++}(s)$ to be a smooth
  function in the whole range $[(\meta-\mpi)^2,(\meta+\mpi)^2]$ is
  that $\re[t_\pm(s)] > t(u_0,s)$ when $s=(\meta-\mpi)^2$. This is
  satisfied when $\qd < q^2_{max}$, $q^2_{max}=(4m^2_\pip-\meta
  m_\piz)(1-m_\piz/m_\eta)=0.051$ GeV$^2$.}.
When $q^2$ increases the zero is no longer present in the first Riemann sheet.

\begin{figure}
\centering
\includegraphics[width=0.6\linewidth]{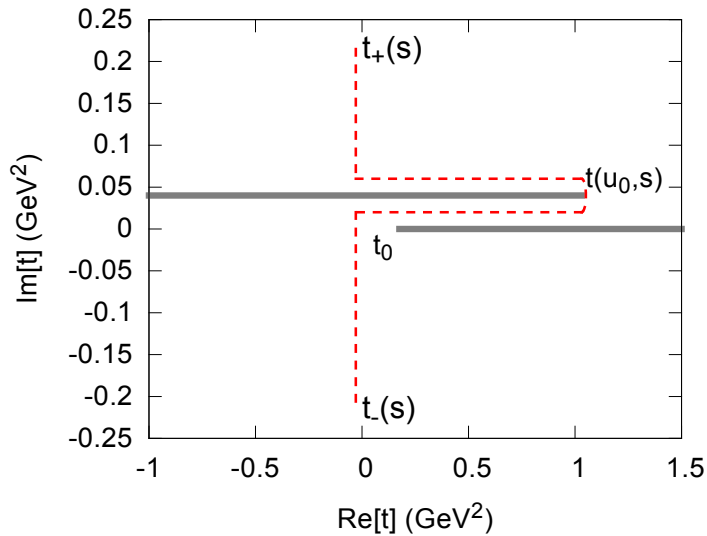}
\caption{\small Illustration of the two cuts of the helicity amplitude $L_{++}(s,t)$
  in the $t$ variable for a fixed value of $s$ in the range $(\meta-\mpi)^2 <
  s < (\meta+\mpi)^2$ (here $s=0.25$ $\hbox{GeV}^2$, $q^2=m_\phi^2$)  and of the
  integration path for performing the partial-wave projection. }
\lblfig{zintpath}
\end{figure}
\subsection{Partial-wave dispersion relations}
One can estimate the asymptotic behaviour of the amplitudes
$L_{\lambda\lambda'}(s,t)$ or $K_{\lambda\lambda'}(s,t)$ when $s\to
\infty$ while $t$ (or $u$) remains finite, based on Regge theory. 
It is also possible to make estimates in the regime where all three
Mandelstam variables go to infinity based on quark counting
rules~\cite{Brodsky:1981rp}. These arguments suggest that the $J=0$
partial waves $l_{0++}(s)$ and $k_{0++}(s)$ should remain bounded by
$\sqrt{s}$ asymptotically and should thus satisfy  once-subtracted
dispersion relations. These should be constrained to satisfy  Low's
soft photon theorem~\cite{Low:1958sn},
\be\lbl{sofphotonrels}
\ba{l}
\left. l_{0++}(s)\right\vert_{s\to\qd}= O(s-\qd)\\[0.2cm]
\left.k^1_{0++}(s)-k^{1,Born}_{0++}(s)\right\vert_{s\to\qd}= O(s-\qd)
\ea\en
where $k^{1,Born}_{0++}$ is the $I=1$, $J=0$ projection of the $K^\pm$ pole
contribution to $K^c_{++}(s,t)$. As we will review below (in
sec.~\sect{bornamplit}) the $S$-wave amplitude $k^{1,Born}_{0++}$ has the
following form 
\be
k^{1,Born}_{0++}(s)= \alpha_B\frac{8\mkpd I_\Kp(s)-2\qd}
{s-\qd},\quad \alpha_B=-\frac{e\,g_{\phi{K}K}}{\sqrt2}
\en
where, for real values of $s$, the function $I_K(s)$ is given by
\be\lbl{IKdetail}
\ba{ll}
I_K(s)=\dfrac{1}{\sqrt{{4\mkd}/{s}-1}}\arctan\sqrt{\dfrac{4\mkd}{s}-1} &
\ :\ 0< s \le 4\mkd \\[0.3cm]
I_K(s)=\dfrac{1}{\sqrt{1-{4\mkd}/{s}}}\arctanh\sqrt{1-\dfrac{4\mkd}{s}} &
\ :\ s<0,\ s >4\mkd
\ea\en
The function $I_K$ is an analytic function of $s$ with a cut along the
negative real axis, 
\be
\im I_K(s+i\epsilon)=\frac{\pi}{2\sqrt{1-4m^2_{\Kp}/s}}\,\theta(-s)\ .
\en 
The amplitude $k^{1,Born}_{0++}(s)$ is thus also an analytic function
of $s$ except for a cut on the negative axis and a pole at $s=\qd$
(which corresponds to the soft photon limit). It satisfies the
following dispersive representation
\be\lbl{dispBorn}
\ba{ll}
k_{0++}^{1,Born}(s)& =\alpha_B\Big[
\dfrac{\beta(\qd)}{s-\qd}+\gamma(\qd) \\ [0.3cm]
\ &+(s-\qd){\displaystyle \int_{-\infty}^0 }\dfrac{4\mkpd }
{ (s'-s)(s'-\qd)^2\sqrt{1-4\mkpd/s'}}     ds'\Big]
\ea\en
with 
\be\lbl{alphaB}
\beta(\qd)=8\mkpd{I_\Kp(\qd)}-2\qd,\quad 
\gamma(\qd)=8\mkpd{I'_\Kp(\qd)}\ .
\en

The $\phi\to\gamma{K}\Kbar$ amplitudes are symmetric in the $t$, $u$
variables due to $C$-invariance.  Singularities in the crossed
channels $\phi{K}\to\gamma{K}$, apart from the charged kaon pole
$\phi{\Kp}\to\Kp\to\gamma{\Kp}$, are induced by unitarity
contributions from $I=1/2$  states: $\pi{K}$, $\pi\pi{K},\,\cdots$. We
will approximate these by the $K^*(892)$ resonance contribution:
$\phi{K}\to K^*\to \gamma{K}$.  These cross-channel singularities
generate a cut in the partial-wave amplitude, denoted as ${\cal C}_{KK}$
illustrated in fig.~\fig{threecuts} below. This cut includes the
negative real axis and also has a complex component.

The amplitudes $\phi\to\gamma\pi\eta$ are not symmetric in the
variables $t$,$u$. This leads to two different cuts in the
partial waves which we will denote as ${\cal C}_{\pi\eta}$ and ${\cal
  C}_{\eta\pi}$ (see fig.~\fig{threecuts}). The cut ${\cal
  C}_{\pi\eta}$ corresponds to unitarity contributions in the
$t$-channel $\phi\eta\to \gamma\pi$ which has isospin $I=0$ i.e. they
are generated from $3\pi$, $5\pi$, $K\Kbar$, $\ldots$ intermediate
states.  We will approximate these by the $\omega$ and $\phi$
resonance contributions: $\phi\eta \to \omega,\phi \to \pi\gamma$.  In
the $u$-channel $\phi\pi\to \gamma\eta$ the unitarity contributions
have $I=1$ states ( $2\pi$, $4\pi$, $K\Kbar$, $\ldots$) and will be
approximated by the $\rho$-meson contribution: $\phi\pi\to \rho \to
\eta\gamma$.  In addition to these left-hand (and complex) cuts the
partial wave amplitudes have a unitarity right-hand cut on the
positive real axis $s > (m_\pi+m_\eta)^2$.

On must also check for the possible presence of an anomalous threshold. 
Anomalous thresholds occur when one of the endpoints of the
left-hand cut requires a deformation of the unitarity
cut~\cite{Mandelstam:1960zz} (see~\cite{Barton:1965} for a review). This
happens for amplitudes involving two virtual photons
(e.g. $\gamma^*\gamma^*\to\pi\pi$,
see~\cite{Hoferichter:2013ama,Hoferichter:2019nlq}). In our 
case, only one photon is virtual. With the $t$, $u$-channel exchanges which we
have considered and the two-body channels included in the unitarity relation
we have checked that no anomalous threshold gets generated when varying the
virtuality $q^2$. 

Finally, taking the soft-photon constraints and the asymptotic bounds
into account one can write the following dispersive integral
representations for the $J=0$ partial waves $l_{0++}$, $k^1_{0++}$,
\be\lbl{dispgeneral} 
l_{0++}(s)=  (s-\qd)\Big[  \dfrac{1}{\pi}
{\displaystyle\int_{{\cal C}_{\pi\eta}+{\cal C}_{\eta\pi}} } 
\dfrac{\disc[l_{0++}(z)]}{(z-s)(z-\qd)}\,dz 
+ \dfrac{1}{\pi}{\displaystyle\int_{m_+^2}^\infty}
\dfrac{\disc[l_{0++}(s')]}{(s'-s)(s'-\qd)}\,ds' \Big]
\en
and
\be\lbl{dispgeneralb}
\ba{ll}
k^1_{0++}(s)=& \alpha_B\Big[\dfrac{\beta(\qd)}{s-\qd}
+\gamma(\qd)\Big]\\[0.4cm]
\           & +(s-\qd)\Big[  \dfrac{1}{\pi}
{\displaystyle\int_{{\cal C}_{KK}}} 
\dfrac{\disc[k^1_{0++}(z)]}{(z-s)(z-\qd)}\,dz 
+ \dfrac{1}{\pi}{\displaystyle\int_{m_+^2}^\infty}
\dfrac{\disc[k^1_{0++}(s')]}{(s'-s)(s'-\qd)}\,ds' \Big]
\ea\en
with 
\be
m_\pm\equiv\meta\pm\mpi\ . 
\en
The discontinuity along the real axis is defined as
\be
\disc[l_{0++}](s')\equiv \frac{1}{2i}\big(l_{0++}(s'+i\epsilon)
-l_{0++}(s'-i\epsilon)\big)\ .
\en
Along the complex contour, the corresponding definition is given in
terms of the parametric representation in eq.~\rf{defdisccx} below.

The dispersive integral representations given in
eqs.~\rf{dispgeneral}~\rf{dispgeneralb} can be considered as exact. In
practice, one intends to estimate the partial-wave amplitudes in the energy
region relevant for the $\phi$ decay i.e. $(\meta+\mpi)^2\le s \le
m^2_\phi$. The dispersive equations~\rf{dispgeneral}~\rf{dispgeneralb} must be
used in an effective theory sense: one tries to perform accurate
approximations to the integrands in a limited energy region e.g. $|s'|\lapprox
1.5$ $\hbox{GeV}^2$ while absorbing the remaining higher energy contributions
into a finite number of low energy parameters.  In this work, we implement the
following approximations:
\begin{itemize}
\item[{\bf a)}] Along the left-hand and complex cuts we approximate
  the discontinuities by those of the light vector mesons ($\rho$,
  $\omega$, $\phi$, $K^*$) and by the $K^+$ exchange contributions to
  the amplitudes, 
\be\lbl{discLCapprox}
\ba{l}
\disc[l_{0++}(s)]_{LC}\approx
{\displaystyle\sum_{V=\rho,\omega,\phi} }\disc[l^V_{0++}(s)]\\[0.4cm]
\disc[k^1_{0++}(s)]_{LC}\approx \disc[k^{1,K^*}_{0++}(s)] +
\disc[k^{1,Born}_{0++}(s)]
\ea\en  
\item[{\bf b)}] Along the right-hand cut we evaluate the discontinuity
  from the unitarity relations, including two channels
\be\lbl{unitrelamplit}
\disc\bp
l_{0++}(s)\\
k^1_{0++}(s)\\
\ep_{RC}= 
\bm{T}^*(s)\bm{\Sigma}(s)
\bp
l_{0++}(s)\\
k^1_{0++}(s)\\
\ep\en
where $\bm{T}(s)$ is the two-channel $\pi\eta-K\Kbar$ $S$-wave
$T$-matrix with
\be
\bm{\Sigma}(s)=\bp
\sigma_{\pi\eta}(s)& 0\\[0.2cm]
0&\sigma_{K}(s)\\
\ep\en
and
\be\lbl{defsigma}
\sigma_{\pi\eta}(s)=\dfrac{\lambda_{\pi\eta}(s)}{s}\theta(s-m_+^2),\quad
\sigma_{K}(s)=\sqrt{\dfrac{s-4\mkd}{s}}\theta(s-4\mkd)\ .
\en
\end{itemize}
Using these approximations for the discontinuities in
eqs.~\rf{dispgeneral}~\rf{dispgeneralb} generates a closed set of
coupled-channel Muskhelishvili equations~\cite{Muskhelishvili}. The
solutions of these equations will include subtraction parameters which
account for the contributions of the higher energy regions in the
dispersive integrals where the approximations made above are no longer
valid.

\subsection{Coupled-channel Omn\`es-Muskhelishvili
  representation}~\lblsec{ccomnes} 
The fundamental ingredient for expressing the solutions of these
equations is the $\bm{\Omega}$ matrix~\cite{Muskhelishvili}. The
matrix elements $\Omega_{ij}(s)$ are defined to be real-analytic
functions of $s$, having a cut on the positive real axis $s\in
[m_+^2,\infty]$ and the discontinuity across this cut being given by
\be\lbl{unitrelomeg}
\im\bm{\Omega}(s)= \bm{T}^*(s)\bm{\Sigma}(s)\bm{\Omega}(s)\ .
\en
In order to write dispersion relations for the matrix elements
$\Omega_{ij}(s)$ we make the assumption that $\Omega_{ij}(s)\sim1/s$ 
when $s\to\infty$.  Other choices can obviously be
made\footnote{In non-relativistic scattering theory, for instance, the
  corresponding Jost matrix is defined  such as to go to the identity
  when $s\to\infty$\cite{Newton:2002}} 
but this one is convenient because it reproduces the asymptotic
behaviour of scalar form-factors in QCD. The $I=1$ form factors are
then related to the matrix elements $\Omega_{ij}$ by simple linear
relations (see~\cite{Albaladejo:2015aca} and
ref.~\cite{Donoghue:1990xh} for the analogous $I=0$ case). Writing
unsubtracted dispersion representations for the $\Omega_{ij}$
functions provides a relation between $\bm{\Omega}(s)$ and the
$S$-wave $T$-matrix in the form of a singular integral matrix equation
\be\lbl{omegintegreq}
\bm{\Omega}(s)=\frac{1}{\pi}\int_{m_+^2}^\infty \frac{ds'}{s'-s}
\bm{T}^*(s)\bm{\Sigma}(s)\bm{\Omega}(s)\ .
\en
In general, this equation must be solved numerically but the determinant
of $\bm{\Omega}$ can be expressed analytically in terms of the two
phase shifts $\delta_{\pi\eta}$, $\delta_{K\Kbar}$,
\be\lbl{detomeg}
\det\bm{\Omega}(s)=\exp\bigg[
\frac{s}{\pi}\int_{m_+^2}^\infty ds'\,
\frac{\delta_{\pi\eta}(s') +\delta_{K\Kbar}(s')}{s'(s'-s)}\bigg]\ .
\en
The phases are defined as continuous functions and the
following asymptotic condition is imposed
\be\lbl{asyphases}
\delta_{\pi\eta}(\infty)+\delta_{K\Kbar}(\infty)=N\pi
\en
where $N$ is the so-called Noether
index~\cite{FritzNoether,Muskhelishvili}. Taking $N=2$ ensures that
eq.~\rf{omegintegreq} has a unique solution with
$\bm{\Omega}(0)=\bm{1}$.  One also sees from eq.~\rf{detomeg} that the
determinant of the Omn\`es matrix does not vanish (except at infinity)
such that the inverse of $\bm{\Omega}$ is defined for all $s$. We will
use the (conventional~\cite{Chew:1960iv}) notation for this inverse
\be
\bm{D}(s)\equiv \bm{\Omega}^{-1}(s)\ .
\en
As a simple consequence of the unitarity relations satisfied by the amplitudes
$l_{0++}(s)$, $k^1_{0++}(s)$ and by the matrix $\bm{\Omega}(s)$
eqs.~\rf{unitrelamplit},~\rf{unitrelomeg}, the two functions obtained by
multiplying the amplitudes by the inverse of the Omn\`es matrix,
have no discontinuity across $[m_+^2,\infty]$,
\be\lbl{Dmultamplit}
\disc\Big[\bm{D}(s)\bp l_{0++}(s)\\
k^1_{0++}(s)\\ 
\ep\Big]_{RC}=0\ .
\en
Based on this property, one can write various types of Omn\`es
representations involving either only right-cut
integrations~\cite{Omnes:1958hv} or only left-cut
integrations~\cite{Frazer:1959gy} or both, which are
exactly equivalent. We consider here a representation starting from
the following two functions,
\be\lbl{phi1phi2def}
\bp
\phi_1(s)\\[0.3cm]
\phi_2(s)
\ep\equiv \bm{D}(s)\bp
l_{0++}(s)       \\[0.3cm]
k^1_{0++}(s)
-\alpha_B\Big[\dfrac{\beta(\qd)}{s-\qd}+\gamma(\qd)\Big]\\
\ep\ .
\en
The terms subtracted from $k^1_{0++}(s)$ remove the soft-photon singularity at
$s=\qd$ such that the functions $\phi_1$, $\phi_2$ vanish at
the soft photon point. These functions have both left and right-hand cuts. 

When $s\to\infty$ the matrix elements $D_{ij}(s)\sim s$ while the
amplitudes $l_{0++}(s)$, $k^1_{0++}(s)$ are expected to grow no faster
than $\sqrt{s}$ (see~\cite{Lu:2020qeo}).  One can thus express
$\phi_1(s)$, $\phi_2(s)$ as twice-subtracted
dispersion relations. It is natural to take $s=\qd$ as one of the
subtraction points while the second one, $s_0$, can be chosen arbitrarily. 
A variation of $s_0$ is compensated by a variation of the
subtraction constants $a_1$, $a_2$. It is convenient to
choose $s_0$ to lie on the real axis and away from the cuts, in the
subthreshold region i.e. in the range $[m_-^2,m_+^2]$. In this case we
can expect the subtraction constants to be essentially real.
The dispersive representations of $\phi_1$, $\phi_2$ involve both
left-cut and right-cut integrations. They can be written as follows,
taking the soft photon constraints into account
\be\lbl{phi1phi2disp}
\bp\phi_1(s)\\[0.2cm]
\phi_2(s)\ep=(s-\qd)\bp
a_1 + I_1^{LC}(s,\qd) + I_1^{RC}(s,\qd) \\[0.2cm]
a_2 + I_2^{LC}(s,\qd) + I_2^{RC}(s,\qd) 
\ep\en
where $a_1$, $a_2$ are subtraction constants and the integrals along the cuts
have the following form
\be\ba{l}
I_i^{LC}(s,\qd)=\dfrac{s-s_0}{\pi}{\displaystyle\int_{{\cal C}_L} }
\dfrac{dz}{(z-s_0)(z-\qd)(z-s)}\disc[\phi_i(z)]\\[0.4cm]
I_i^{RC}(s,\qd)=\dfrac{s-s_0}{\pi}{\displaystyle\int_{m_+^2}^\infty } 
\dfrac{ds'}{(s'-s_0)(s'-\qd)(s'-s)} \disc[\phi_i(s')]\ .
\ea\en
Let us now express these integrals in more detail.
\begin{itemize}
\item[{\bf 1)}]{\bf Right-cut integrals:} 
The discontinuities of the functions $\phi_i$ along the right-hand cut are
driven by the singular parts of the Born amplitude. Their expressions are
deduced from~\rf{phi1phi2def} using~\rf{Dmultamplit}
\be
\disc\bp \phi_1(s)\\
\phi_2(s)\\ \ep_{RC}= -\alpha_B
\Big[ \frac{\beta(\qd)}{s-\qd}+\gamma(\qd)\Big]\,\disc\bp
D_{12}(s)\\
D_{22}(s)\\ \ep\ .
\en
Using the analyticity properties of the matrix elements $D_{ij}(s)$
these right-cut integrals $I^{RC}_i(s,\qd)$ can be calculated explicitly
and one obtains
\be\lbl{IRC_1}
I^{RC}_i(s,\qd)=
-\alpha_B\left( I^{RC,\beta}_i(s,\qd)
+ I_i^{RC,\gamma}(s,\qd)\right)
\en
with
\be\lbl{IRC_2}
\ba{ll}
I^{RC,\beta}_i(s,\qd)=& \beta(\qd)\Big[ \dfrac{1}{s-\qd}
\Big(\dfrac{D_{i2}(s)-D_{i2}(\qd)}{s-\qd}-D'_{i2}(\qd)\Big)\\[0.3cm]
& -\dfrac{1}{s_0-\qd}
\Big(\dfrac{D_{i2}(s_0)-D_{i2}(\qd)}{s_0-\qd}-D'_{i2}(\qd)\Big)\Big]\\[0.4cm] 
I^{RC,\gamma}_i(s,\qd)=& \gamma(\qd)\Big[
\dfrac{D_{i2}(s)-D_{i2}(\qd)}{s-\qd}
               -\dfrac{D_{i2}(s_0)-D_{i2}(\qd)}{s_0-\qd}\Big]\ .
\ea\en

\item[{\bf 2)}]{\bf Left-cut integrals:}
Along the left-hand cuts now, the discontinuities of the functions
$\phi_i(s)$ are proportional to those of the amplitudes $l_{0++}$,
$k^1_{0++}$. Using the model defined in eq.~\rf{discLCapprox}, one can
write the integrals $I_i^{LC}$ in eq.~\rf{phi1phi2def} as a sum over
five terms 
\be\lbl{ILC_i}
I_i^{LC}(s,\qd)=I_i^{LC,Born}(s,\qd)
+\sum_{V=\rho,\omega,\phi,K^*}  I_i^{LC,V}(s,\qd)\ .
\en
The integrals $I_i^{LC,Born}$  are associated with the
left-cut discontinuity induced by the kaon pole, they read 
\be
I_i^{LC,Born}(s,\qd)=\alpha_B(s-s_0)\int_{-\infty}^0 ds'\frac{4\mkpd\,D_{i2}(s')}
{(s'-s_0)(s'-\qd)^2(s'-s)\sqrt{1-4\mkpd/s')}}\ .
\en
One remarks that the double pole at $s'=\qd$ in the integrand does not cause any
numerical difficulty when $\qd >0$ and that the integrals are real when $s$ is
in the physical region (also assuming that the subtraction point
$s_0$ is real and positive). 

The transitions $\phi\to \omega\eta$, $\phi\to \phi\eta$ or $\phi\to K^*\Kbar,
\bar{K}^* K$ are all kinematically forbidden. In those cases, ignoring the width
of the vector resonances is a good approximation. 
In the zero-width limit, the expressions for the integrals
$I_i^{LC,V}$appearing in~\rf{ILC_i} are as follows, using the results
from the next section on the discontinuities of the vector-exchange
amplitudes,
\be\lbl{ILC,V}
\ba{l}
I_i^{LC,\omega}(s,\qd)=\alpha_\omega(s-s_0){\displaystyle\int_{{\cal C}_{\pi\eta}}}
\dfrac{dz\, z\,D_{i1}(z)\,sgn(z)}
{\lambda_{\pi\eta}(z)(z-s_0)(z-s)} \Psi_{\omega\pi}(z,\qd)\\[0.5cm]
I_i^{LC,\phi}(s,\qd)=\alpha_\phi(s-s_0){\displaystyle\int_{{\cal C}_{\pi\eta}}}
\dfrac{dz\, z\,D_{i1}(z)\,sgn(z)}
{\lambda_{\pi\eta}(z)(z-s_0)(z-s)} \Psi_{\phi\pi}(z,\qd)\\[0.5cm]
I_i^{LC,{K^*}}(s,\qd)=
\alpha_{K^*}(s-s_0){\displaystyle\int_{{\cal C}_{KK}}}
\dfrac{dz\, z\,D_{i2}(z)\,sgn(z)}
{\lambda_{KK}(z)(z-s_0)(z-s)} \Psi_{K^*K}(z,\qd)
\ea\en
where
\be
\Psi_{VP}(z,\qd)=\frac{1}{z-\qd}\left(-m_V^2+\qd\left(\frac{m_V^2-m_P^2}
{z-\qd}\right)^2\right) 
\en
and $sgn(z)$ is a sign factor, $sgn(z)=\pm1$ (see sec.~\sect{dispvex}). 
The parameters  $\alpha_\omega$, $\alpha_\phi$, $\alpha_{K^*}$ 
are related to resonance chiral Lagrangian coupling constants introduced
in the next section, 
\be\lbl{alpha_V}
\ba{l}
\alpha_\omega=\dfrac{e C_{\omega\pi\gamma} g_{\phi\omega\eta}}{2}\\[0.2cm]
\alpha_\phi={e C_{\phi\pi\gamma} g_{\phi\phi\eta}}  \\[0.2cm]
\alpha_{K^*}\equiv -\dfrac{e}{\sqrt2}\left(
 C_{K^{*+}\Kp\gamma} g_{\phi{K^{*+}\Km}}
-C_{K^{*0}\Kz\gamma} g_{\phi{K^{*0}\Kzb}}  \right)\ .
\ea\en

In the case of the $\rho$ meson, the effect of the width must be taken
into account because the transition $\phi\to \rho \pi$ is
kinematically allowed.
This can be done in a simple way, consistently with the
analyticity properties, by replacing the $u$-channel pole $1/(u-m_\rho^2)$ by
a K\"allen-Lehmann dispersive
representation~\cite{Kallen:1952zz,Lehmann:1954xi}. Denoting the zero-width
limit of the $\rho$ integrals as $I_i^{LC,\rho}(s,\qd;m_\rho)$ with
\be\lbl{ILC,rho0}
I_i^{LC,\rho}(s,\qd;m_\rho)=
\alpha_\rho(s-s_0){\displaystyle\int_{{\cal C}_{\eta\pi}}}
\dfrac{dz\, z\,D_{i1}(z)\,sgn(z)}
{(z-s_0)(z-s)\lambda_{\pi\eta}(z)}\Psi_{\rho\eta}(z,\qd)
\en
where
\be\lbl{alpha_rho}
\alpha_\rho=\dfrac{e C_{\rho\eta\gamma} g_{\phi\rho\pi}}{2}\ ,
\en
one can express the finite-width result as an integral involving the
corresponding K\"allen-Lehmann spectral function
$\rho^V(\mu^2;m_\rho,\Gamma_\rho)$,
\be\lbl{ILC,rho1}
I_i^{LC,\rho}(s,\qd;m_\rho,\Gamma_\rho)=
\int_{4\mpid}^\infty d\mu^2\, \rho^V(\mu^2;m_\rho,\Gamma_\rho)
I_i^{LC,\rho}(s,\qd;\mu)
\en 
We will show in sec.~\sect{finitewidth} how to simplify this expression
using integration by parts. 
\end{itemize} 
The radiative decay amplitudes $V\to \gamma P_1P_2$ have often been
expressed in a form which involves the charged kaon (or charged pion)
one-loop triangle function(e.g.~\cite{Achasov:1987ts,Nussinov:1989gs,LucioMartinez:1990uw,Close:1992ay,Bramon:1992ki,Oller:1998ia,Marco:1999df,Bramon:2000vu}),
which seems different from the Omn\`es
representations presented above. We show in appendix~\sect{kaonloopderiv} that
expressions in terms of the triangle loop function can indeed be
derived from our dispersive representation for certain
specific models of $T$-matrices which, in particular, have no
left-hand cut. Physically, of course, all the $T$-matrix elements
for $\pi\eta-K\Kbar$ scattering do have left-hand cuts (and a complex
cut in the case of $\pi\eta\to \pi\eta$) such that the Kaon-loop
representation is not exactly valid.

\section{Born and vector-exchange amplitudes}
In this section we review the evaluation of the Born amplitude 
and that of the vector-exchange amplitudes 
based on a resonance chiral Lagrangian.
The Born amplitude is proportional to the $\phi\to \Kp\Km$ coupling
constant, the magnitude of which can be evaluated from experiment.
The vector meson exchange contributions to $\phi\to \gamma
K\Kbar$ and $\phi\to \gamma \piz\eta$ involve the product of a
radiative decay $V\to \gamma P$ coupling, the magnitude of
which can be determined from experiment, and a hadronic $V_1
\to V_2 P$ coupling.  Among those, only the $\omega\to\rho\pi$ and $\phi\to
\rho\pi$ couplings can be estimated from experiment. We can use
flavour symmetry and large $N_c$ arguments in order to derive
estimates for the other $V_1\to V_2 P$ couplings which are
needed. This will also allow us to relate the sign of the Born
amplitude to those of the vector exchange amplitudes.

\subsection{Resonance chiral Lagrangian and mixing angle}
We start from the following resonance chiral Lagrangian~\cite{Ecker:1989yg}
\be\lbl{LagVchir}
\ba{ll}
{\cal L}^V= & -\dfrac{1}{4}\braque{V_{\mu\nu} V^{\mu\nu}} 
+\dfrac{1}{2} M_V^2 \braque{V_\mu  V^\mu}
-\dfrac{1}{2\sqrt2} f_V \braque{V_{\mu\nu} f_+^{\mu\nu}}\\[0.3cm]
\  &-\dfrac{ig_V}{2\sqrt2}\braque{V_{\mu\nu} [u^\mu,u^\nu]}
+\epsilon_{\mu\nu\alpha\beta}\Big\{
h_V \braque{ V^\mu\{u^\nu,f_+^{\alpha\beta}\} }\\[0.3cm] 
\  &+\dfrac{1}{2}\sigma_V \braque{V^\mu\{u^\nu,V^{\alpha\beta}\}} 
+i\theta_V\braque{V^\mu u^\nu u^\alpha u^\beta} \Big\}
\ea\en
with $V_{\mu\nu}=\nabla_\mu V_\nu-\nabla_\nu V_\mu$.  We differ from
ref.~\cite{Ecker:1989yg} in considering a nonet (rather than an octet)
of vector mesons, which are encoded in a $3\times3$ matrix with
\be
V_{\mu}=\frac{1}{\sqrt2}\sum_{a=0}^8 V^a_{\mu} \lambda_a\ .
\en
The Lagrangian~\rf{LagVchir} is of leading order in the large $N_c$ and in the
chiral expansions i.e. $O(N_c)$, $O(m_q^0)$ .  We consider only the couplings
which are relevant to the amplitudes of interest here. In the terms
proportional to $\epsilon_{\mu\nu\alpha\beta}$ we use the same notation
as~\cite{Prades:1993ys}. The pseudoscalar fields $P_a$ are encoded, as usual,
in a unitary matrix $U=\exp(i\lambda_a P_a/F_\pi)$ and $u_\mu\equiv u^\dagger
D_\mu U u^\dagger$ with $u=\sqrt{U}$.  The external vector and axial-vector
sources can be set to $v_\mu= eQ A_\mu$, $a_\mu=0$, so that
\be
D_\mu U=\partial_\mu U-ie [Q,U] A_\mu, \quad
f_+^{\mu\nu}= e\,(uQu^\dagger+u^\dagger Q u)F^{\mu\nu}
\en
where $A_\mu$ is the photon field. With these definitions we can
express the matrix element of the electromagnetic current,
\be\lbl{f_V}
\braque{0\vert j^{em}_\mu(x)\vert V^a(p_V,\lambda)}= -e m^2_V f_V 
  \braque{\lambda_aQ} e^{-ip_V x}\,e_\mu(\lambda)\ .
\en
In order to get the correct pattern for the masses of the vector mesons we
must add mass terms which break both the flavour and the
nonet symmetries,
\be\lbl{LagVsb}
{\cal L}^V_{sb}=
\frac{1}{2}\lambda \braque{{\cal M}_q V_\mu V^\mu}
+ \frac{1}{2}\epsilon_V V^0_\mu V^{0\mu} \ 
\en
where ${\cal M}_q$ is the quark mass matrix and the parameter $\epsilon_V$ is
sub-leading in the large $N_c$ expansion i.e. $O(N_c^0)$.
These flavour and nonet symmetry-breaking terms
induce a mixing between the $\omega$ and the $\phi$ mesons, with a mixing
angle $\varphi_V$, such that the $\lambda$ matrices attached to the
physical $\omega$ and $\phi$ mesons can be written as
\be
\lambda_\omega=\bp
c_1&0&0\\
0  &c_1&0\\
0&0&\sqrt2{s_1}\\
\ep,\quad  
\lambda_\phi=\bp
s_1&0&0\\
0&s_1&0\\
0&0&-\sqrt2{c_1}\\
\ep\en
with
\be
c_1\equiv\cos(\varphi_1),\quad s_1\equiv\sin(\varphi_1)\ ,\quad
\varphi_1=\varphi_I-\varphi_V
\en
where $\varphi_I=\arctan(1/\sqrt2)\simeq 35.26^\circ$ is the ``ideal'' mixing
angle. We can first derive two relations between the Lagrangian parameters and
the $\rho$ and $K^*$ meson masses and then, the masses of the $\omega$ and
$\phi$ mesons are obtained by diagonalising the singlet-octet mass
matrix. After a rotation by the ideal mixing angle this matrix reads
\be
\tilde{{\cal M}}=\bp
m^2_\rho+\frac{2}{3}\epsilon_V & -\frac{\sqrt2}{3} \epsilon_V \\[0.3cm]
-\frac{\sqrt2}{3} \epsilon_V & 2m^2_{K^*}-m^2_\rho+ \frac{1}{3}\epsilon_V
\ep\ .
\en
In this form, it is easy to see that
the experimental fact that $m_\rho\simeq m_\omega$ implies that
$\epsilon_V << m^2_\rho,m^2_{K^*}$. To leading order in $\epsilon_V$, one
obtains for the angle $\varphi_1$
\be\lbl{teta1epsV}
\varphi_1= -\frac{\sqrt2 \epsilon_V}{6 (m_{K^*}^2-m_\rho^2)}\ .
\en
This formula shows that $\varphi_1$, even
though numerically small, is actually chirally enhanced because the
denominator is $O(m_s-m_{ud})$. This provides some justification for
neglecting further Lagrangian terms which break the nonet symmetry.
The masses of the $\omega$ and
$\phi$ mesons (i.e. the eigenvalues of $\tilde{\cal M}$) are given by
\be\lbl{momeg+mphi}
m_\omega^2=m_\rho^2 +\frac{2}{3}\epsilon_V,\quad
m_\phi^2=2m_{K^*}^2-m_\rho^2 +\frac{1}{3}\epsilon_V
\en
to leading order in $\epsilon_V$. This implies the mass relation
\be\lbl{massrel}
2m^2_\phi-m^2_\omega= 4m^2_{K^*}-3m^2_\rho\ .
\en 
which is verified within 4\%. 
This small discrepancy, however, gives rise to a significant relative
uncertainty in the evaluation of $\varphi_1$.
Using the $\omega$ mass gives $\varphi_1=-1.2^\circ$ while using the
$\phi$ mass gives $\varphi_1=-7.3^\circ$. 
\subsection{Signs of the coupling constants}
One can derive useful information on the signs of the coupling
constants by using asymptotic constraints on the $V\to P\gamma^*$ form
factors.  Let us define the form-factor $F_{VP}$ as follows
\be\lbl{F_VPdef}
\braque{V(p_V,\lambda)\vert j_\mu^{em}(0)\vert P(p_P)}=
2e C_{VP\gamma}\epsilon_{\mu\nu\alpha\beta}p_P^\nu p_V^\alpha
e^{*\beta}_V(\lambda)\, F_{VP}(q^2)
\en
where $q=p_V-p_P$. We have factored out the coupling $C_{VP\gamma}$
such that the form factor must satisfy $F_{VP}(0)=1$, which ensures
that the amplitude $V\to P\gamma^*$ becomes equal to $V\to P\gamma$
when $q^2=0$. Computing the form factor in the flavour symmetry limit
from the Lagrangian~\rf{LagVchir} one obtains~\cite{Prades:1993ys}
\be
F_{V^aP^b}(q^2)=1 + \dfrac{\sigma_V f_V}{\sqrt2 h_V}\dfrac{ q^2}{M_V^2-q^2}
\en
Requiring that the form factor goes to zero when $q^2$ goes to
infinity gives
\be
\sigma_V f_V =\sqrt2 h_V\ .
\en
Combining this result with the relation
\be\lbl{fv=2gv}
f_V\simeq 2g_V
\en
which can be derived from a similar asymptotic constraint on the $\gamma\pi$
axial form factor~\cite{Ecker:1989yg} one obtains
\be\lbl{sigmaVhV}
\sigma_V h_V\simeq \frac{h^2_V}{\sqrt2\, g_V}\ .
\en 
This relation fixes the relative signs between the vector-exchange
amplitudes which are proportional to $\sigma_V h_V$ and the Born
amplitude which is proportional to $g_V$. Without loss of generality
we can choose $g_V$, $h_V$ and $\sigma_V$ to be positive.

\subsection{$VP\gamma$ and $VVP$ coupling constants}
We will need a set of  $VP\gamma$ and $VVP$ couplings which can be
defined from the Lagrangians
\be
\ba{l}
{\cal L}_{VP\gamma}= \epsilon_{\mu\nu\alpha\beta}\,
{\displaystyle\sum_{a,b}} e\,C_{V_aP_b\gamma}\,  
V_a^\mu\,\partial^\nu{P_b}\,F^{\alpha\beta}\\[0.3cm]
{\cal L}_{\phi VP}= \epsilon_{\mu\nu\alpha\beta}\,
{\displaystyle\sum_{a,c}} g_{\phi V_a P_c}\, 
\partial^\alpha{\phi^{\beta}}\, V_a^\mu\,\partial^\nu{P_c}
\ea\en
The magnitude of the $VP\gamma$ couplings which are needed can all be
determined from the experimental values of the radiative decay widths,
\be
\Gamma_{V_a\to P_b\gamma}= \alpha \vert C_{V_aP_b\gamma}\vert^2
\frac{(m^2_{V_a}-m^2_{P_b})^3}{6 m^3_{V_a}}\ .
\en
The results are collected in table~\Table{CVPgamma} which also shows
the expressions of these couplings using the resonance chiral chiral
Lagrangian~\rf{LagVchir} with the leading order nonet symmetry breaking
terms~\rf{LagVsb}. For the amplitudes which involve an $\eta$ meson we
use a simple $\eta-\eta'$  mixing description such that the $\lambda$
matrix attached to the $\eta$ meson is
\be
\lambda_\eta=\bp 
s_\eta& 0 & 0\\[0.2cm]
0& s_\eta&  0\\[0.2cm]
0& 0  &-\sqrt{2}c_\eta\\
\ep,\quad
s_\eta=\sin(\varphi_I-\varphi_P),\ 
c_\eta=\cos(\varphi_I-\varphi_P)
\en
and take $\varphi_P=-20^\circ$. 
These expressions allow us to determine the signs
of the $C_{VP\gamma}$ couplings as shown in the table.

\begin{table}
\centering \bt{c|c|c||c|c|c}\hline\hline 
\TT $C_{VP\gamma}$ & flavour &
exp. (GeV$^{-1})$ & $g_{VVP}$ &flavour & exp.(GeV$^{-1})$\\ \hline
$C_{K^{*+}K^+\gamma}$ & $ -\dfrac{2\sqrt2\, h_V}{3\fpi}$ & $-0.412(21)$ &
$g_{\phi K^{*+}K^+}$ & $\dfrac{2(\sqrt2 c_1-s_1)\,\sigma_V}{\fpi}$ &$\simeq+10.0$
\\[0.3cm] 
$C_{K^{*0}K^0\gamma}$ & $ \dfrac{4\sqrt2\, h_V}{3\fpi}$ &  $+0.635(27)$ & 
$g_{\phi K^{*0}K^0}$ & $\dfrac{2(\sqrt2 c_1-s_1)\,\sigma_V}{\fpi}$ &$\simeq+10.0$ 
\\[0.3cm] 
$C_{\omega\piz\gamma}$ & $-\dfrac{2\sqrt2 c_1 h_V}{\fpi}$ & $-1.157(16)$ & 
$g_{\phi\omega\eta}$   &$-\dfrac{4 c_1 s_1(\sqrt2 c_\eta+s_\eta)\sigma_V}{\fpi}$ 
& $\simeq+1.4$ 
\\[0.3cm]
$C_{\phi\piz\gamma}$ & $ -\dfrac{2\sqrt2 s_1 h_V}{\fpi}$ & $+0.067(1)$ &
$g_{\phi\phi\eta}$ &$\dfrac{2(\sqrt2 c_1^2c_\eta-s_1^2s_\eta)\,\sigma_V}{\fpi}$
& $\simeq+5.4$  
\\[0.3cm] 
$C_{\rho^0\eta\gamma}$ &$ -\dfrac{2\sqrt2 s_\eta h_V}{\fpi}$ &$-0.790(28)$ & 
$g_{\phi\rho^0\piz}$ &$-\dfrac{4s_1\,\sigma_V}{\fpi}$ & $+0.80\pm 0.15$ 
\\[0.3cm] \hline 
\BB $C_{\rho^0\pi^0\gamma}$ &$ -\dfrac{2\sqrt2 h_V}{3\fpi}$ & $-0.368(23)$ &
$g_{\omega\rho^0\piz}$   &$-\dfrac{4c_1\,\sigma_V}{\fpi}$ & $-(14.8\pm3.0)$ \\
\hline\hline 
\et 
\caption{\small  $VP\gamma$ and $VVP$ coupling constants needed for computing the
  vector-exchange amplitudes (see text).}\ .
\lbltab{CVPgamma} 
\end{table}

Let us now consider the $g_{VVP}$ couplings. As is well known, one can
estimate the two couplings $g_{\omega\rho\pi}$, $g_{\phi\rho\pi}$ from the
experimental values of the $\omega, \phi \to 3\pi$ decay widths. The
$V\to3\pi$ decay amplitude derived from the resonance chiral
Lagrangian~\rf{LagVchir} can be written in the following form
\be\lbl{Vto3pi}
{\cal T}_{V(\lambda)\to \pip\pim\pi0}=\frac{ 2 g_{V\rho\pi} g_V}{\fpid}
\epsilon(e_V(\lambda),p_1,p_2,p_3) 
({\cal F}(s) + {\cal F}(t) + {\cal F}(u))
\en
with
\be
{\cal F}(z) =\frac{\theta_V}{2\sqrt2 g_V \sigma_V} -z P_\rho(z)\ 
\en
where $P_\rho(z)$ is the $\rho$-meson propagator taking the width into account
(see sec.~\sect{dispvex}). Imposing the following asymptotic
constraint\footnote{This can be justified by matching to the asymptotic
  behaviour in the Brodsky-Lepage regime\cite{Lepage:1980fj}: $s\to\infty$,
  $s/t$ fixed.}
 on ${\cal F}$
\be
\left.{\cal F}(z)\right\vert_{z\to\infty}= O(1/z) 
\en
and using that $P_\rho(z)\sim 1/z$ when $z\to\infty$ yields the following
relation which determines the value of $\theta_V$
\be\lbl{thetaVrel}
\theta_V= 2\sqrt2 g_V \sigma_V\ .
\en
The amplitude~\rf{Vto3pi} then becomes identical to that of the original GSW
model~\cite{GellMann:1962jt} and to the one obtained from effective
Lagrangians implementing a hidden-gauge symmetry~\cite{Fujiwara:1984mp}.  
From the expression of the  $V\to \pip\pim\piz$ width which reads
\be
\ba{ll}
\Gamma_{V\to \pip\pim\piz}= & \dfrac{1}{3 (4\pi)^3 m_V}
\left(\dfrac{g_{V\rho\pi} g_V}{\fpid}  \right)^2 \\[0.4cm]
\ &
\times{\displaystyle \int_{4m^2_\pip}^{(m_V-m_\piz)^2} }ds
{\displaystyle \int_{\tau^-(s)}^{\tau^+(s)}} dt\, 
\vert \vec{p}_1\wedge \vec{p}_2\vert^2
\vert {\cal F}(s)+ {\cal F}(t)+ {\cal F}(u)\vert^2
\ea\en
with
\be
\tau^\pm(s)=\frac{1}{2}(m^2_V+2m^2_\pip+m^2_\piz -s \pm
\sigma_\pip(s)\lambda_{V \piz}(s) )
\en
one obtains the following values for the $g_{V\rho\pi}$ couplings
\be\lbl{g_Vrhopiexp}
\vert g_{\phi\rho\pi}\vert=0.80\pm 0.01,\quad
\vert g_{\omega\rho\pi}\vert=14.8\pm 0.1\ 
\ (\hbox{GeV}^{-1}).
\en
The value of $g_{\omega\rho\pi}$ is compatible with the result derived
from the experimental measurements of the form factor
$F_{\omega\pi}(s)$ in the region $s > (m_\omega+m_\pi)^2$ in
ref~\cite{Achasov:2016zvn}. The errors quoted in eq.~\rf{g_Vrhopiexp}
do not contain the uncertainties induced by the modelling. Varying
$\theta_V$ from the value given by eq.~\rf{thetaVrel} by 20\% induces
a variation of $g_{\omega\rho\pi}$ by 14\% and a variation of
$g_{\phi\rho\pi}$ by 8\%. It must also be kept in mind that these
simple modellings of the $V\to 3\pi$ amplitudes and of the $F_{V\pi}$
form factors do not correctly account for the unitarity relations and
the related dispersive representations~\cite{Niecknig:2012sj,
Schneider:2012ez,Danilkin:2014cra,Albaladejo:2020smb}. We finally
ascribe a 20\% uncertainty to the values of $g_{\omega\rho\pi}$ and
$g_{\phi\rho\pi}$. Using these two inputs, together with the
$C_{VP\gamma}$ couplings as shown in the table one can determine the values of
the parameters $\sigma_V$, $h_V$ and $\theta_1$ from a least-squares fit. This
yields 
\be
\ba{l}
{h_V}     = (0.38\pm0.04)\, \fpi,\\[0.2cm]
{\sigma_V}= (3.39\pm0.47)\, \fpi,\\[0.2cm]
\theta_1=-(3.60\pm 0.08)^\circ\ .
\ea\en
One can then deduce estimates for the values of the couplings 
$g_{\phi K^*K}$, $g_{\phi\omega\eta}$ and $g_{\phi\phi\eta}$ which are
needed in order to evaluate all the relevant vector-exchange amplitudes, 
The corresponding numerical values of the effective couplings
  $\alpha_V$ which appear in front of the $I=1$ partial-wave
amplitudes (see~\rf{alpha_V}~\rf{alpha_rho}) are given by
\be\lbl{alpha_Vnumer}
\ba{ll}
\alpha_\rho  & =-(0.096\pm 0.018)\ \hbox{GeV}^{-2}\\
\alpha_\omega& \simeq-0.245\ \hbox{GeV}^{-2}\\
\alpha_\phi  & \simeq+0.110\ \hbox{GeV}^{-2}\\
\alpha_{K^*} & \simeq+2.242\ \hbox{GeV}^{-2}\ .
\ea\en 
\begin{figure}
\centering
\includegraphics[width=0.48\linewidth]{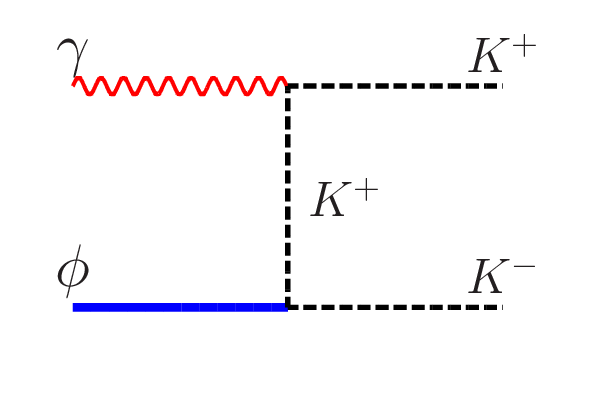}
\caption{\small Illustration of the charged kaon exchange (Born)
  contribution to the $\phi\to \gamma \Kp\Km$ amplitude (the crossed
  diagram is not shown).}  
\lblfig{phi_born}
\end{figure}
\subsection{Born  amplitude}\lblsec{bornamplit}
The Born amplitude in $\phi\to\gamma\Kp\Km$ (see
fig.~\fig{phi_born}) can now be computed from the
Lagrangian~\rf{LagVchir}. Including the 
contributions proportional to the two couplings $g_V$ and  $f_V$ gives
\be\lbl{bornPhi}
K_{Born}^{\mu\nu}= -e\frac{ g_V (\sqrt2c_1+s_1)m^2_\phi} {2\fpid}
\frac{1}{(t-\mkd)(u-\mkd)} \left(-2q_1\cdot q_2\, T_1^{\mu\nu}
+\frac{1}{2} T_2^{\mu\nu}\right) 
+{\cal T}_a^{\mu\nu} \ ,
\en
with
\be\lbl{polyTa}
{\cal T}_a^{\mu\nu}=\frac{e (-2g_V+f_V) (\sqrt2c_1+s_1)}{2\fpid}T_1^{\mu\nu}\ .
\en
This contribution is suppressed because of the approximate relation
$f_V\simeq 2g_V$. Actually,
our Omn\`es representations use only the singular parts of the tree
amplitudes, other contributions being absorbed into the subtraction
constants. Terms like~\rf{polyTa} which make a constant contribution
to the $S$-wave can thus be omitted.  
We can replace $g_V$ in eq.~\rf{bornPhi}  using the
relation with the $\phi\Kp\Km$ coupling 
\be
g_{\phi\Kp\Km}=\frac{g_V(\sqrt2c_1+s_1)m^2_\phi}{2\fpid}\ .
\en
The coupling $g_{\phi\Kp\Km}$ can be determined from the
$\phi\to\Kp\Km$ decay width, 
\be
\Gamma_{\phi\to\Kp\Km}= \frac{g^2_{\phi\Kp\Km}}{48\pi} m_\phi\left(
1-\frac{4m^2_\Kp}{m^2_\phi}\right)^{3/2}
\en
which gives, 
\be
g_{\phi\Kp\Km}=4.48\pm0.02\ ,
\en
recalling that $g_V$ was chosen to have a positive sign. 
The three independent Born helicity amplitudes deduced from
eq.~\rf{bornPhi} read
\be
\ba{l}
K^{Born}_{++}=-\dfrac{e g_{\phi KK}}{\qd-s}
\left(\dfrac{8\mkpd}{1-(1- 4\mkpd/s)\cos^2\theta}-2\qd\right)\\[0.4cm]
K^{Born}_{+-}=-\dfrac{e g_{\phi KK}}{\qd-s}\,
\dfrac{2(s-4\mkpd)\sin^2\theta}{1-(1- 4\mkpd/s)\cos^2\theta}\\[0.4cm]
K^{Born}_{+0}=-\dfrac{e g_{\phi KK}}{\qd-s}\sqrt{\dfrac{\qd}{2s}}
\dfrac{4(s-4\mkpd)\sin\theta\cos\theta}{1-(1- 4\mkpd/s)\cos^2\theta}\  .
\ea\en
The  $J=0$ partial-wave amplitude is then easily obtained 
\be
k^{Born}_{0++}(s,\qd)=\frac{e\,g_{\phi KK}}{s-\qd}\left(
8\mkpd I_\Kp(s)-2\qd\right)
\en
where $I_\Kp(s)$ was given in eq.~\rf{IKdetail}.
\begin{figure}
\centering
\includegraphics[width=0.48\linewidth]{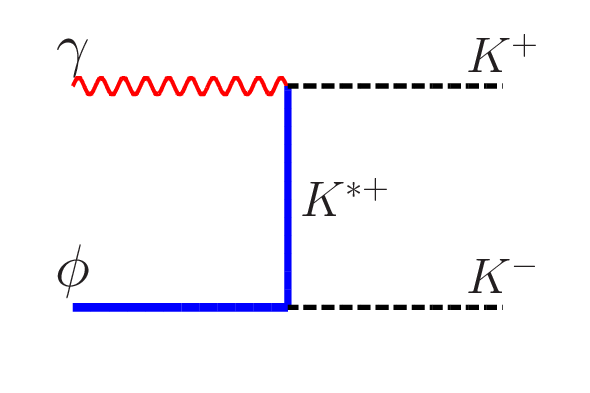}\includegraphics[width=0.48\linewidth]{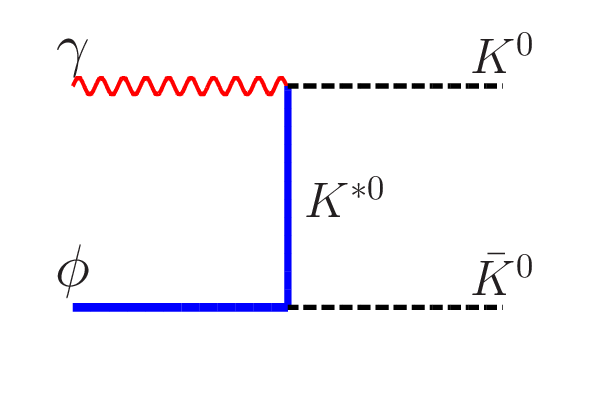}\\
\includegraphics[width=0.48\linewidth]{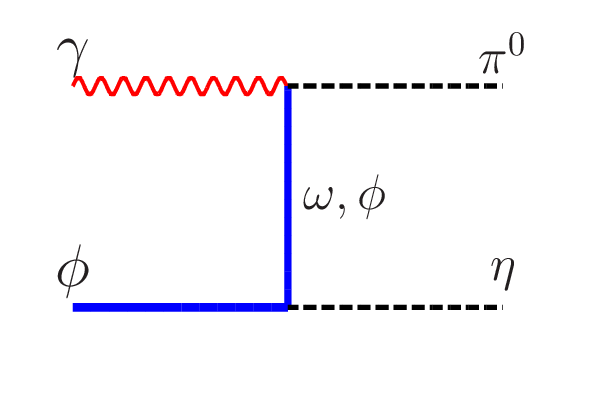}\includegraphics[width=0.48\linewidth]{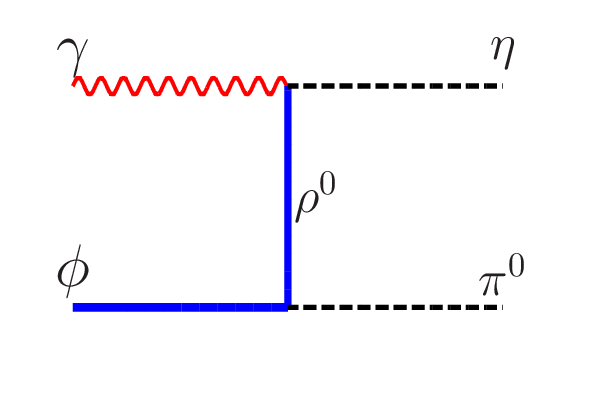}\\
\caption{\small Illustration of the vector meson exchange
  contributions to $\phi\to \gamma K\Kbar$ and $\phi\to \gamma \pi
  \eta$ amplitudes considered here. }
\lblfig{phi_vect}
\end{figure}

\subsection{Vector-exchange amplitudes}
We can now express the vector-exchange amplitudes (see
fig.~\fig{phi_vect}) in terms of the couplings
listed in table~\Table{CVPgamma}. These amplitudes involve the
following two combinations of the three independent tensors $T_i^{\mu\nu}$
\be
\Sigma_\pm^{\mu\nu}=
-2(q_1\cdot q_2 +\Delta^2\pm (q_1+q_2)\cdot\Delta)T_1^{\mu\nu} 
+\frac{1}{2}T_2^{\mu\nu} \pm T_3^{\mu\nu}
\en
The $K^*$ exchange contributions to $\phi\to \gamma \Kp\Km, \gamma \Kz\Kzb$ read
\be\lbl{Ktensor}
\ba{l}
{K}^{\mu\nu}_{K^{*+}}= 
\dfrac{e\, C_{K^{*+}K^+\gamma} g_{\phi K^{*+}K^-}}{4}\left(
 \dfrac{1}{t-m^2_{K^{*+}}} \Sigma_+^{\mu\nu}
+\dfrac{1}{u-m^2_{K^{*+}}} \Sigma_-^{\mu\nu} \right)\\[0.3cm]
{K}^{\mu\nu}_{K^{*0}}= 
\dfrac{e\, C_{K^{*0}K^0\gamma} g_{\phi K^{*0}K^0}}{4}\left(
\dfrac{1}{t-m^2_{K^{*0}}} \Sigma_+^{\mu\nu}
+\dfrac{1}{u-m^2_{K^{*0}}} \Sigma_-^{\mu\nu} \right)\ 
\ea
\en
and the contributions from $\omega$, $\phi$ and $\rho$ exchanges to 
$\phi\to \gamma\piz \eta$ read
\be\lbl{Ltensor}
\ba{l}
{L}^{\mu\nu}_\omega= 
\dfrac{e\, C_{\omega\pi\gamma} g_{\phi\omega\eta}}{4}\,
\dfrac{1}{t- m^2_\omega} \Sigma_+^{\mu\nu}\\[0.4cm]
{L}^{\mu\nu}_\phi= 
\dfrac{e\, C_{\phi\pi\gamma} g_{\phi\phi\eta}}{2}\,
\dfrac{1}{t- m^2_\phi}\Sigma_+^{\mu\nu} \\[0.4cm]
{L}^{\mu\nu}_\rho= 
\dfrac{e\, C_{\rho\eta\gamma} g_{\phi\rho\pi}}{4}\,
\dfrac{1}{u- m^2_\rho}\Sigma_-^{\mu\nu}\ .
\ea
\en
These expressions are derived in the zero-width limit. Finite width
effects will be discussed in sec.~\sect{dispvex}.

The helicity amplitudes corresponding  to eqs.~\rf{Ktensor},~\rf{Ltensor}
are deduced using the relations~\rf{helicABC} and the
partial-wave amplitudes are  then easily computed. The $J=0$
partial-wave projections can be expressed in terms of the
following generic function ${\cal F}_V$
\be\lbl{genericF}
\ba{l}
{\cal F}_V(s,\qd,\mvd,m_1^2,m_2^2)\equiv\Big[-\mvd +\qd
  \Big(\dfrac{\mvd-m_1^2}{\qd-s}\Big)^2\Big]
  \,L_V(s,\qd,\mvd,m_1^2,m_2^2)\\[0.3cm] 
\quad\quad
+\dfrac{\qd}{2}
\Big(1+\dfrac{m_1^2-m_2^2}{s}+2\,\dfrac{\mvd-m_1^2}{\qd-s}\Big)+s-\qd
\ea\en
with
\be\lbl{LVdef}
\ba{l}
L_V(s,\qd,\mvd,m_1^2,m_2^2)=\\[0.2cm]
\quad\quad\dfrac{s}{\lambda_{12}(s)}\big[\log(\mvd-t_+(s,\qd,m_1^2,m_2^2))
-\log(\mvd-t_-(s,\qd,m_1^2,m_2^2))\big]
\ea
\en
and
\be
t_\pm(s,\qd,m_1^2,m_2^2)=m_1^2+\frac{\qd-s}{2s}
   (s+m_1^2-m_2^2\pm \lambda_{12}(s))\ . 
\en
The expressions of the $\phi\to\gamma K\Kbar$ and $\phi\to\gamma
\pi\eta$ $S$-wave amplitudes associated with vector meson exchanges
are given below in terms of the ${\cal F}_V$ functions,
\be\lbl{k_0++}
\ba{ll}
k_{0++}^{K^{*+}}(s)=& 
      {e\, C_{K^{*+}K^+\gamma} g_{\phi K^{*+}K^-}}\,
      {\cal F}_V(s,\qd,m_{K^{*+}},m_{K^+},m_{K^+})\\[0.3cm]
k_{0++}^{K^{*0}}(s)=& 
      {e\, C_{K^{*0}K^0\gamma} g_{\phi K^{*0}K^0}}\,
      {\cal F}_V(s,\qd,m_{K^{*0}},m_{K^0},m_{K^0})\\[0.3cm]
\ea\en
and
\be\lbl{l_0++V}
\ba{ll}
l_{0++}^{\omega}(s)=& 
      \dfrac{e\, C_{\omega\pi\gamma} g_{\phi\omega\eta}}{2}\,
      {\cal F}_V(s,\qd,m^2_\omega,m^2_\pi,\metad)\\[0.3cm]
l_{0++}^{\phi}(s)=& 
       {e\, C_{\phi\pi\gamma} g_{\phi\phi\eta}}\,
      {\cal F}_V(s,\qd,m^2_\phi,m^2_\pi,\metad)\\[0.3cm]
l_{0++}^{\rho}(s)=& 
      \dfrac{e\, C_{\rho\eta\gamma} g_{\phi\rho\pi}}{2}\,
      {\cal F}_V(s,\qd,m^2_\rho,\metad,m^2_\pi)\ .
\ea\en

\section{Integrations along the complex cuts}

\subsection{Parametric representations of the cuts}
It is easy to check that the function ${\cal F}_V$ vanishes in the limit $s\to
\qd$, as it should. Concerning the singularities, ${\cal F}_V$ has a pole at
$s=0$ when the two masses $m_1$, $m_2$ are different. The function 
$\lambda_{12}(s)$ has square-root singularities at $s=(m_1\pm m_2)^2$ but
these are not present in ${\cal F}_V$ because it is an even function of
$\lambda_{12}$.  The remaining singularity of ${\cal F}_V$ is a cut generated
by the log functions. This cut is given by the values of $s$ such that the
argument of one of the log's is a negative real number, that is,
\be\lbl{tpm=t}
t_\pm(s,\qd,m_1^2,m_2^2)=t,\quad t \ge \mvd\ .
\en
Solving for $s$ as a function of $t$ provides the following parametric
representation of the cut,
\be\lbl{paramcut}
s_\pm(t,\qd,m_1^2,m_2^2)= \qd -\frac{t-m_1^2}{2t}\left(\qd +t -m_2^2
\pm \sqrt{\lambda(t,\qd,m_2^2)} \right)\ .
\en
An important point is that the shape of the cut does not depend on the mass of
the exchanged vector meson. Only the lower bound on the variable $t$ depends
on this mass, see~\rf{tpm=t}. The cut of the $\phi\to\gamma \pi\eta$
amplitudes has two components, denoted as ${\cal C}_{\pi\eta}$ and ${\cal
  C}_{\eta\pi}$ corresponding to eq.~\rf{paramcut} with $m_1,m_2 =m_\pi,
m_\eta$ and $m_\eta, m_\pi$ respectively.  The left-cut of the $\phi\to\gamma
K\Kbar$ amplitudes, denoted as ${\cal C}_{KK}$ is given by eq.~\rf{paramcut}
with $m_1=m_2=m_K$. The cuts lie on the real axis when $t < (q-m_2)^2$ or $t >
(q+m_2)^2$ otherwise they are complex. One sees that $s_+$ goes to $-\infty$
when $t\to \infty$. The $s_-$ function can be re-written as
\be
s_-(t,\qd,m_1^2,m_2^2)= \qd\Big(1-\frac{2(t-m_1^2)}{\qd +t -m_2^2
+\sqrt{\lambda(t,\qd,m_2^2)}}\Big)
\en
which shows that $s_-$ goes to 0 when $t\to\infty$. We also note that
in the kinematical configuration where $q \ge m_V+m_2$ the endpoints
of the cut $s_\pm(m_V,\qd,m_1^2,m_2^2)$ are located on top of the
unitarity cut. This corresponds to the triangle diagram singularities
interpreted in ref.~\cite{Coleman:1965xm}.

\begin{figure}
\centering \includegraphics[width=0.48\linewidth]{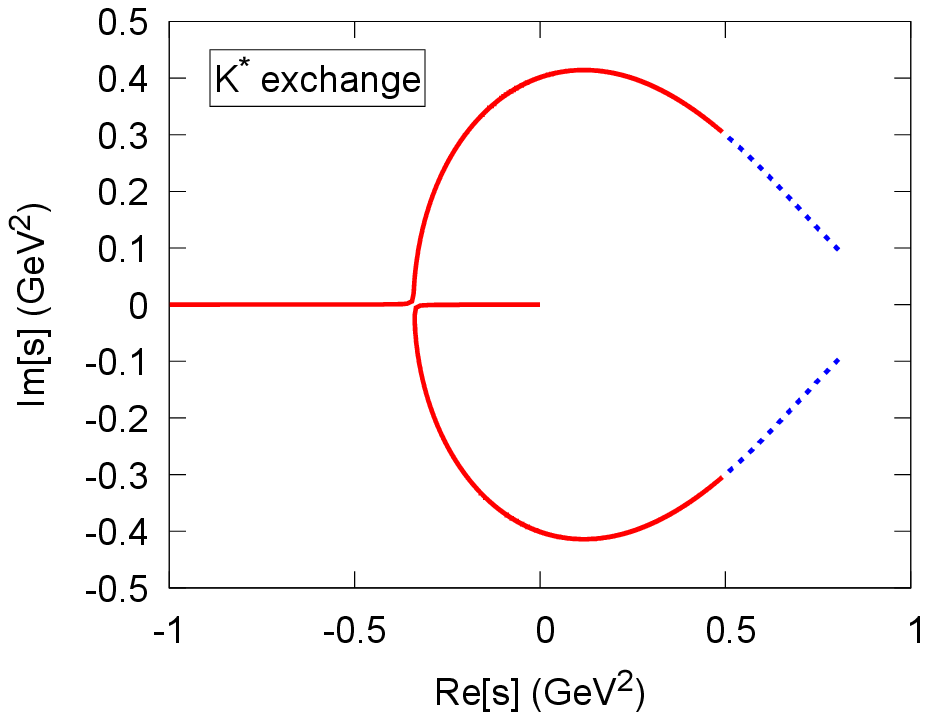}%
\includegraphics[width=0.48\linewidth]{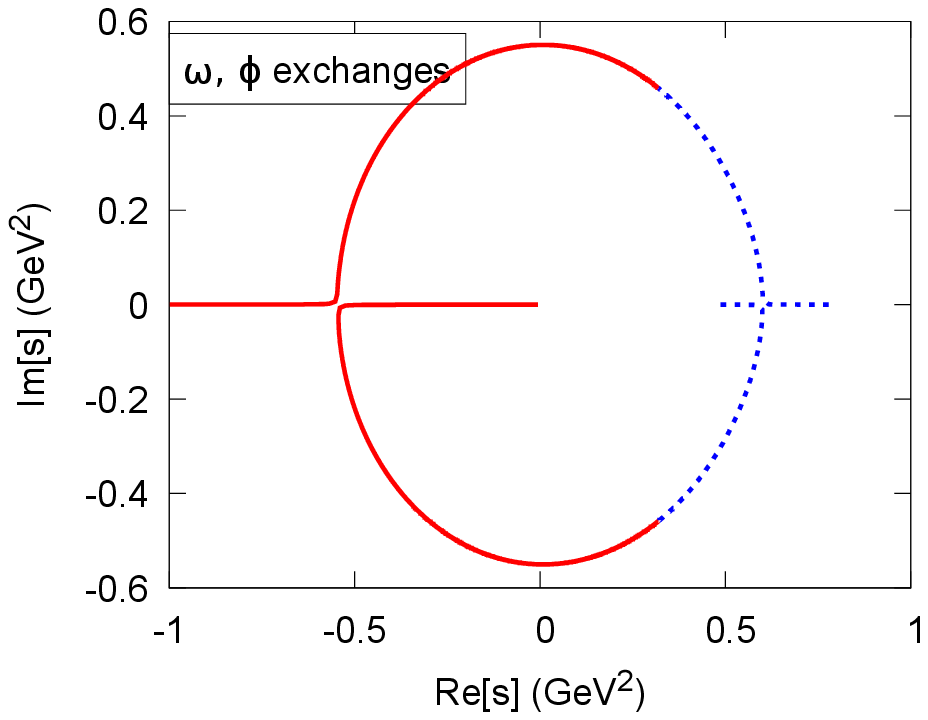}\\[-0.4cm]
\includegraphics[width=0.48\linewidth]{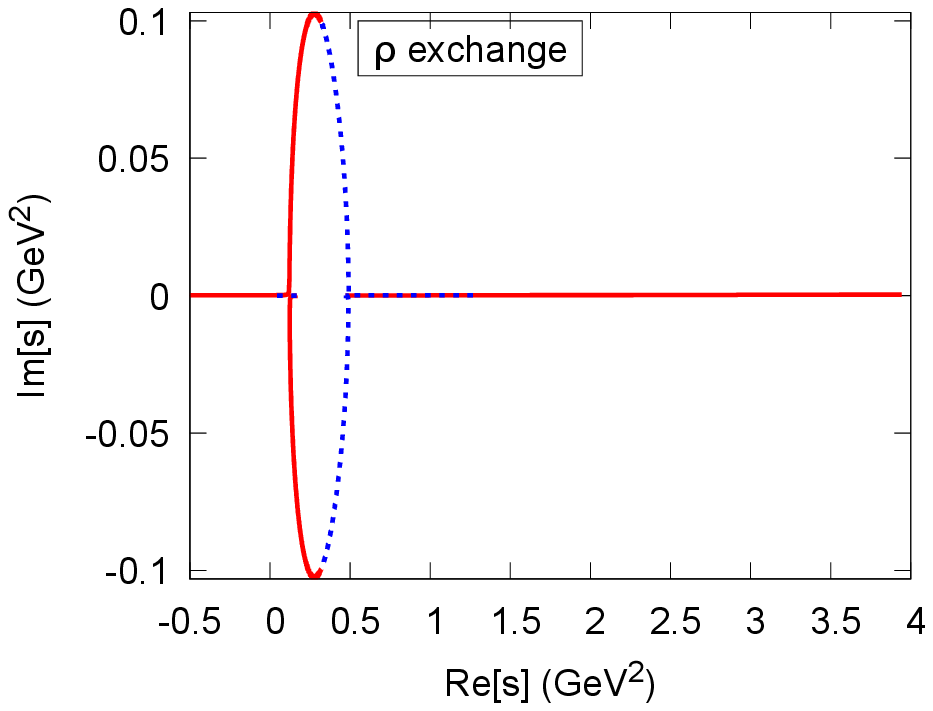}%
\includegraphics[width=0.48\linewidth]{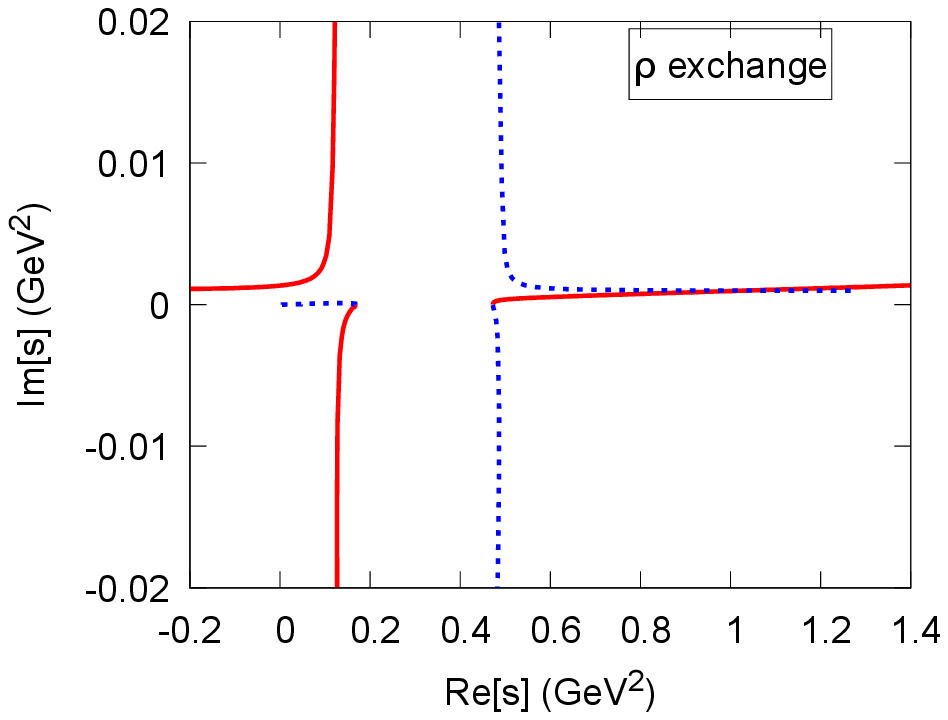}
\caption{\small Complex cuts of the partial-wave
  amplitudes $\phi\to \gamma K\Kbar$ and $\phi\to \gamma\pi\eta$. The portions
  in dotted blue (solid red) correspond to $sgn(z)=-1$ ($=+1$), where $sgn(z)$
  is the sign factor, see~\rf{DeltaF}. In the figures, the lower bound on the
  parameter $t$ in the representation~\rf{paramcut} has been taken as
  $t_0=(m_K+m_\pi)^2$ for the $K^*$ exchange, $t_0=9m^2_\pi$ for $\omega$
  exchange and $t_0=4m^2_\pi$ for $\rho$ exchange. The effect of the
  $\qd+i\epsilon$ prescription is illustrated in the right bottom plot.}
\lblfig{threecuts}
\end{figure}
\subsection{Testing the dispersive representations on the vector-exchange
  amplitudes}\lblsec{dispvex} 
We first consider the $J=0$ partial-wave projections of the vector-exchange
amplitudes and discuss their dispersive
representations as integrals over the complex cuts. This allows us
to check the correctness and the numerical accuracy of such integrals
before extending them to the complete Omn\`es representation of the $S$-wave.

The generic function ${\cal F}_V$ which appears in the expressions of
the $S$-waves (see~\rf{genericF}) must satisfy the
following Cauchy representation as an integral over the cut $C_{12}$,
\be\lbl{cauchycalF}
{\cal F}_V(s,\qd,\mvd,m_1^2,m_2^2)= 
(s-\qd)\Big\{1 -\frac{m_1^2-m_2^2}{2s}+I_{12}(s,\qd)
\Big\}
\en
with
\be
I_{12}(s,\qd)=\frac{1}{\pi}\int_{{\cal C}_{12}}
\dfrac{dz}{(z-\qd)(z-s)}\,\disc[{\cal F}_V(z,\qd)] 
\en
taking into account the asymptotic behaviour, the pole\footnote{As
  discussed in sec.~\sect{pwsingular} the  singularity at $s=0$ in the full
  amplitude is weaker than a pole.}  at $s=0$ and the
cut. The discontinuity across the cut is defined as follows
\be\lbl{defdisccx}
\disc[{\cal F}_V( s_\pm(t),\qd)]=\lim_{\epsilon\to0}
\frac{ {\cal F}_V(s_\pm(t)+i\epsilon \dot{s}_\pm(t),\qd)
-{\cal F}_V(s_\pm(t)-i\epsilon \dot{s}_\pm(t),\qd)}
{2i}\ 
\en
(where $\dot{s}_\pm(t)\equiv d{s}_\pm(t)/dt$).
Using the explicit form of ${\cal F}_V$ (eq.~\rf{genericF}) the discontinuity
reads
\be\lbl{DeltaF}
\left.\disc[{\cal F}_V(z,\qd)]\right\vert_{z\in {\cal C}_{12} }=\pi\Big[-\mvd +\qd
  \Big(\dfrac{\mvd-m_1^2}{\qd-z}\Big)^2\Big]\frac{z}{\lambda_{12}(z)}sgn(z)
\en
where $sgn(z)=\pm 1$, depending on which one of the
two log functions in $L_V$ generates the discontinuity. 
The integrations along the cut are expressed as follows, using the parametric
representations
\be\lbl{intdzparametric}
\int_{{\cal C}_{12}} f(z)\,dz =\int_{\mvd}^\infty  f({s}_+(t)) \,\dot{s}_+(t)dt\
+ \int_{\mvd}^\infty f({s}_-(t))\,\dot{s}_-(t)dt\ .
\en
Fig.~\fig{threecuts} illustrates the three different cuts 
${\cal  C}_{\pi\eta}$, ${\cal C}_{\eta\pi}$, ${\cal C}_{KK}$ which are
involved and shows the regions where the sign factor in the
discontinuity~\rf{DeltaF} $sgn(z)$ is $+1$ or $-1$ in different colour. For
${\cal C}_{KK}$  the sign change occurs when the parameter
$t=q^2-m_K^2$ and for ${\cal  C}_{\pi\eta}$ when $t=q^2-\metad$. Along
the cut  ${\cal C}_{\eta\pi}$ there are several sign changes which are
indicated in table~\Table{signfactors}.
\begin{table}[h]
\centering\bt{c|c|c|c|c}
$t$    & $[4\mpid,t_1]$ & $[t_1,q^2-\mpid]$ &
$[q^2-\mpid,t_2]$ & $[t_2,\infty]$\\ \hline
$sgn(s_+(t))$ & $+$ & $-$ & $+$ & $+$ \\ 
$sgn(s_-(t))$ & $-$ & $-$ & $+$ & $-$ \\ \hline
\et
\caption{\small  Values of the sign factor along the contour 
  ${\cal  C}_{\eta\pi}$ as a function of the parameter $t$, with
$t_1=\meta(q^2/(\meta+\mpi)-\mpi)$, 
$t_2=\meta(q^2/(\meta-\mpi)+\mpi)$ .}
\lbltab{signfactors}
\end{table}

These three cuts include the negative real axis $[-\infty,0]$, as we
have seen, and a complex component. In addition, the $\pi\eta$ cuts
extend along the positive real axis. These cuts can be shifted away
from the unitarity cut by appending an infinitesimally small positive
imaginary part to $\qd$. Nevertheless, the cut ${\cal C}_{\eta\pi}$
crosses the real axis at the two points $m^2_\pm=(\meta\pm
\mpi)^2$. When the energy $s$ crosses one of these points the
discontinuity of the $\pi\eta$ partial wave exhibits a divergence
induced by the factor $1/\lambda_{\eta\pi}(s)$. This is in accordance
with the discussion in sec.~\sect{pwsingular} on the path of
integration for performing the partial-wave projection when $s$ is in
the range $m^2_-<s<m_+^2$.

Vector mesons are resonances and thus have a finite width.  The width
can be simply taken into account, in a way compatible with analyticity
properties, by replacing the pole in the $t$ or $u$ variable in the
tensorial vector-exchange amplitudes~\rf{Ktensor}~\rf{Ltensor} by a
dispersive K\"allen-Lehmann~\cite{Kallen:1952zz,Lehmann:1954xi}
representation
\be\lbl{kallenlehman}
\frac{1}{u-\mvd}\longrightarrow P_\rho(u;m_V,\Gamma_V)=
\int_{t_0}^\infty d\mu^2\,
\frac{\rho^V(\mu^2;m_V,\Gamma_V)}{u-\mu^2}
\en
where $t_0=4m^2_\pi$ in the case of the $\rho$.  We will assume that the
spectral function $\rho^V$ satisfies the normalisation condition
\be\lbl{normalroV}
\int_{t_0}^\infty  d\mu^2\,\rho^V(\mu^2;m_V,\Gamma_V)=1\ 
\en
which ensures that the propagator goes as $1/u$ 
when $u$ goes to infinity. 
In the case of the $\rho$ meson, for instance, we will use the following model
for the spectral function\footnote{This model differs from the one used in
  ref.~\cite{Moussallam:2013una} by a factor $\sqrt{t}$, which gives rise to
  somewhat simpler formulae.}
\be\lbl{rhoVmodel}
\rho^V(\mu^2)=N_V\dfrac{ \gamma_V\,(\mu^2-t_0)^{3/2}}
{ \mu^2\,(\mu^2-m^2_\rho)^2 +\gamma_V^2 (\mu^2-t_0)^3}\theta(\mu^2-t_0)
\en
where $\gamma_V$ is proportional to the physical width $\Gamma_\rho$ 
\be
\gamma_V =\frac{\Gamma_\rho}{m_\rho} \left( \frac{m^2_\rho}
{m^2_\rho-t_0}\right)^{3/2}
\en
and $N_V$ is adjusted such that eq.~\rf{normalroV} is satisfied.  It is easy
to see that the model~\rf{rhoVmodel} satisfies the condition that $\rho^V$
should tend to a $\delta$ function when the width $\Gamma_V$ goes to zero,
\be
\lim_{\Gamma_V\to0} \rho^V(\mu^2;m_V,\Gamma_V)=\delta(\mu^2-\mvd)\ .
\en
Using the finite width propagators~\rf{kallenlehman}, the vector-exchange
partial-wave amplitudes, get expressed as integrals over $\rho^V$, e.g.
\be
l^V_{0++}(s;m_V,\Gamma_V)=\int_{t_0}^\infty d\mu^2 \rho^V(\mu^2;m_V,\Gamma_V)\,
l_{0++}(s;\mu)
\en 
The corresponding dispersive representations have the same form as
eq.~\rf{cauchycalF} in which the functions $I_{12}(s)$ are expressed as
double integrals,
\be\lbl{I12width}
\ba{l@{}l}
I_{12}(s,\qd;m_V,\Gamma_V)=
{\displaystyle\sum_{a=\pm}\int_{t_0}^\infty }d\mu^2\rho^V(\mu^2;m_V,\Gamma_V)
&{\displaystyle\int_{\mu^2}^\infty} dt\, \dfrac{\dot{s}_a(t)\,
  {s}_a(t) sgn_a(t)}
{({s}_a(t)-s)({s}_a(t)-\qd) \lambda_{12}( {s}_a(t) )}\\[0.4cm]
\  &\quad\times\Big[-\mu^2 +\qd\dfrac{ (\mu^2-m_1^2)^2}{(s_a(t)-\qd)^2}
\Big]\ .
\ea\en
These expressions can be simplified using integration by parts. Let us
introduce the following two integrals of $\rho^V$,
\be\lbl{rhoV1rhoV2}
\ba{l}
\rho^V_1(t)\equiv {\displaystyle\int_{m_1^2}^t} d\mu^2\, \mu^2 
\rho^V(\mu^2;m_V,\Gamma_V)\\[0.4cm]
\rho^V_2(t)\equiv {\displaystyle\int_{m_1^2}^t} d\mu^2\, (\mu^2-m_1^2)^2 
\rho^V(\mu^2;m_V,\Gamma_V)
\ea\en
which can be computed analytically using the model~\rf{rhoVmodel} (see
appendix~\sect{roV}). Integrating by parts, the functions $I_{12}$ get
expressed in the following way in terms of $\rho^V_1$ and $\rho^V_2$
\be\lbl{I12singleint}
\ba{l} 
I_{12}(s,\qd;m_V,\Gamma_V)=
\rho^V_1(t_0)\Phi^{(1)}(s,\qd)-\rho^V_2(t_0)\,\qd\Phi^{(3)}(s,\qd)\\[0.4cm]
\quad+{\displaystyle\sum_{a=\pm}\int_{t_0}^\infty }dt\,\Big[
-\rho^V_1(t)+ \rho^V_2(t)\dfrac{\qd}{(s_a(t)-\qd)^2} \Big]
\dfrac{\dot{s}_a(t)\, {s}_a(t) \,sgn_a(t)}
{({s}_a(t)-s)({s}_a(t)-\qd) \lambda_{12}( {s}_a(t) )}\
\ea
\en
which involves a single integration.
The two terms involving $\Phi^{(n)}(s,\qd)$ are boundary contributions
which will be detailed below. A remark is in order here: one sees from
eq.~\rf{paramcut} that $s_+(t)-\qd$, $s_-(t)-\qd$ both vanish when
$t=m_1^2$. This generates a potentially highly singular term in the integral
of eq.~\rf{I12singleint}. This problem only affects the contour ${\cal
  C}_{\eta\pi}$ since, in this case, $m^2_1=\meta^2$ which lies within the
range of integration. The functions $\rho^V_1(t)$, $\rho^V_2(t)$ were chosen
such as to vanish when $t=m_1^2$ (see~\rf{rhoV1rhoV2}) and this removes
completely these singularities in the $t$-integral~\rf{I12singleint}. The
singular integrations are now contained in the two functions
$\Phi^{(1)}(s,\qd)$, $\Phi^{(3)}(s,\qd)$
\be\lbl{Phi_n}
\ba{ll}
\Phi^{(n)}(s,\qd)  &= {\displaystyle\sum_{a=\pm}\int_{t_0}^\infty} dt 
\dfrac{\dot{s}_a(t) s_a(t)\,sgn_a(t)}{\lambda_{12}(s_a(t))\,
(s_a(t)-s)(s_a(t)-\qd)^n}\\[0.45cm]
 &={\displaystyle\int_{{\cal C}_{12}}  }dz\, \dfrac{z\,sgn(z)}
{\lambda_{12}(z) (z-s)(z-\qd)^n}\ .
\ea\en
They can be evaluated analytically in terms of the function
$L_V(s,\qd,\mvd=t_0,m_1^2,m_2^2)$, see~\rf{LVdef} 
\be
\ba{l}
\Phi^{(1)}(s,\qd)= \dfrac{1}{s-\qd} L_V(s,\qd,t_0,m_1^2,m_2^2)\\[0.3cm]
\Phi^{(2)}(s,\qd)= \dfrac{1}{s-\qd}\left(
\Phi^{(1)}(s,\qd)         -\dfrac{1}{t_0-m_1^2}\right)\\[0.3cm]
\Phi^{(3)}(s,\qd)=\dfrac{1}{s-\qd}\left(
\Phi^{(2)}(s,\qd)+\dfrac{\qd+\Delta_{12}}{2\qd(t_0-m_1^2)^2}\right)\ .
\ea\en

\begin{figure}
\centering 
\includegraphics[width=0.60\linewidth]{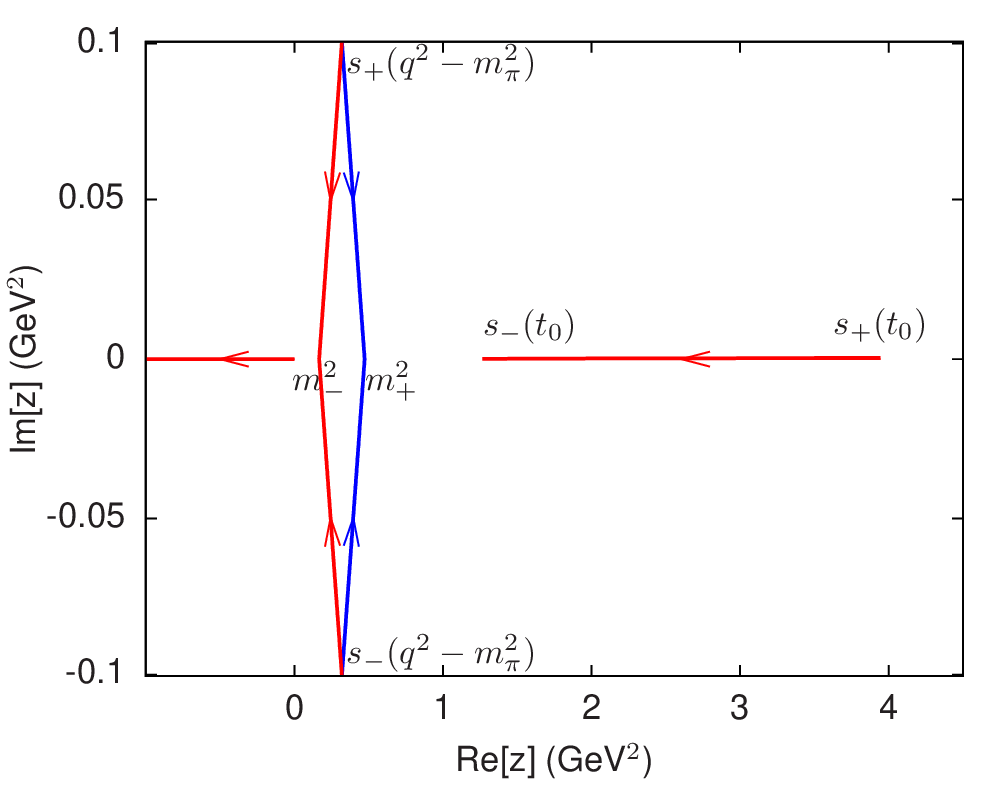}
\caption{\small Deformed contour used for the numerical computation of the
  functions $\bar{\Phi}_i^{(n)}(s,\qd)$ (see text).}
\lblfig{cxcutdeform}
\end{figure}

\subsection{Omn\`es integrations with a finite resonance
  width}\lblsec{finitewidth}     
Let us now return to the Omn\`es representations and reconsider the left-cut
integrals $I_i^{LC,\rho}$, which were given in eq.~\rf{ILC,rho1} in the form
of double integrals.  The discussion in the preceding section can be easily
adapted to derive expressions in the form of single integrals which are easy
and fast to compute. One obtains,
\be\lbl{ILC,rho}
\ba{ll} 
I_i^{LC,\rho}(s,\qd)& =\alpha_\rho(s-s_0)\bigg\{
\rho^V_1(t_0)\bar{\Phi}_i^{(1)}(s,\qd)
-\rho^V_2(t_0)\qd\bar{\Phi}_i^{(3)}(s,\qd)\\[0.4cm]
\ &+{\displaystyle\sum_{a=\pm}\int_{t_0}^\infty }dt\,\bigg[
-\rho^V_1(t)+ \rho^V_2(t)\dfrac{\qd}{(s_a(t)-\qd)^2} \bigg]\\[0.4cm]
\ & \times\dfrac{s_a(t) \dot{s}_a(t)\, D_{i1}({s}_a(t)) \,sgn_a(t)}
{(s_a(t)-s_0)({s}_a(t)-\qd) ({s}_a(t)-s)\lambda_{\pi\eta}( {s}_a(t) )}\bigg\}
\ea
\en
where $s_\pm(t)$ parametrise the contour ${\cal C}_{\eta\pi}$ and the
functions $\bar{\Phi}_i^{(n)}$ are given by
\be\lbl{tildePhi_n}
\bar{\Phi}_i^{(n)}(s,\qd) =
{\displaystyle\int_{{\cal C}_{\eta\pi}}  }dz\,\dfrac{ z\,D_{i1}(z)\,sgn(z)}
{(z-s_0)(z-\qd)^n(z-s)\lambda_{\pi\eta}(z)} \ .
\en
These integrals must be computed numerically. When $s$ is in the physical
range of the $\phi$ decay, $(\meta+\mpi)^2\le s \le \qd$, the numerical
integration can be performed in a way which avoids the singularities at
$z=\qd$ and $z=s$ by deforming the contour.  Taking into account the exact
cancellations in the regions of overlapping branches which have a different
sign factor (see fig.~\fig{threecuts}) the contour can be deformed into a) the
segment $[s_-(t_0),s_+(t_0)]$ on the real axis (note that $q^2 < s_-(t_0)\simeq
1.266\ \hbox{GeV}^2$), b) the four segments starting from $(\meta\pm\mpi)^2$
and joining $s_\pm(\qd-\mpid)$ and c) the negative real axis. This deformed
contour, showing the directions of integrations, is illustrated in
fig.~\fig{cxcutdeform}.

\subsection{Recovering the Adler zero}\lblsec{B-Riemann}
The component ${\cal C}_{\eta\pi}$ of the complex cut in the $l_{0++}$
partial wave which is induced by the $\rho$ meson exchange (and more
generally by $u$-channel exchanges) is responsible for the
disappearance of the Adler zero, as was mentioned above. Indeed, this
cut crosses the real axis at the two points $s=m_+^2$ and $s=m_-^2$
and this causes a divergence of the partial wave when $s$ approaches
$m_+^2$ from below or $m_-^2$ from above. It is possible to recover
the Adler zero by considering an analytical continuation of the
amplitude to an unphysical Riemann sheet which we will call the
$B$ Riemann sheet.  Let us illustrate this in the case of the simple
$\rho$-exchange amplitude $l^\rho_{0++}(s)$. The $B$ Riemann sheet
extension $l^{\rho,B}_{0++}(s)$ is defined such as to match
continuously with $l^\rho_{0++}(s)$ upon crossing the portion of the
cut ${\cal C}_{\eta\pi}$ which extends from $m_+^2$ to $s_+(t_0)$. In
the zero-width limit the two functions are simply related by
\be\lbl{lrhoIIzerow}
\left.l^{\rho,B}_{0++}(s)\right\vert_{\Gamma_\rho=0}
=l^\rho_{0++}(s)+\alpha_\rho \frac{2{\pi}i\, s}{\lambda_{\eta\pi}(s)}
\Big(-m_\rho^2 +  (m_\rho-\metad)^2\frac{\qd }{(s-\qd)^2}\Big)
\en 
This formula is easily generalised to the case of a finite width using the
integral representation of the amplitude and one finds
\be\lbl{lrhoIIfinitew}
l^{\rho,B}_{0++}(s)=l^\rho_{0++}(s)+\alpha_\rho \frac{2{\pi}i\,
  s}{\lambda_{\eta\pi}(s)} 
\Big(-\bar{\rho}^V_1(t_-(s))
+\bar{\rho}^V_2(t_-(s))\frac{\qd }{(s-\qd)^2}\Big)
\en
where the $\bar{\rho}^V_1(t)$, $\bar{\rho}^V_2(t)$ are integrals of the spectral
function which vanish at $t=t_0$ (see~\rf{rhobarVdef}).  These $\rho$-exchange
partial-wave amplitudes are illustrated in fig.~\fig{l0pp_rho} which shows
that the B-sheet extension displays an Adler zero close to $s=\metad$.

\begin{figure}
\centering
\includegraphics[width=0.6\linewidth]{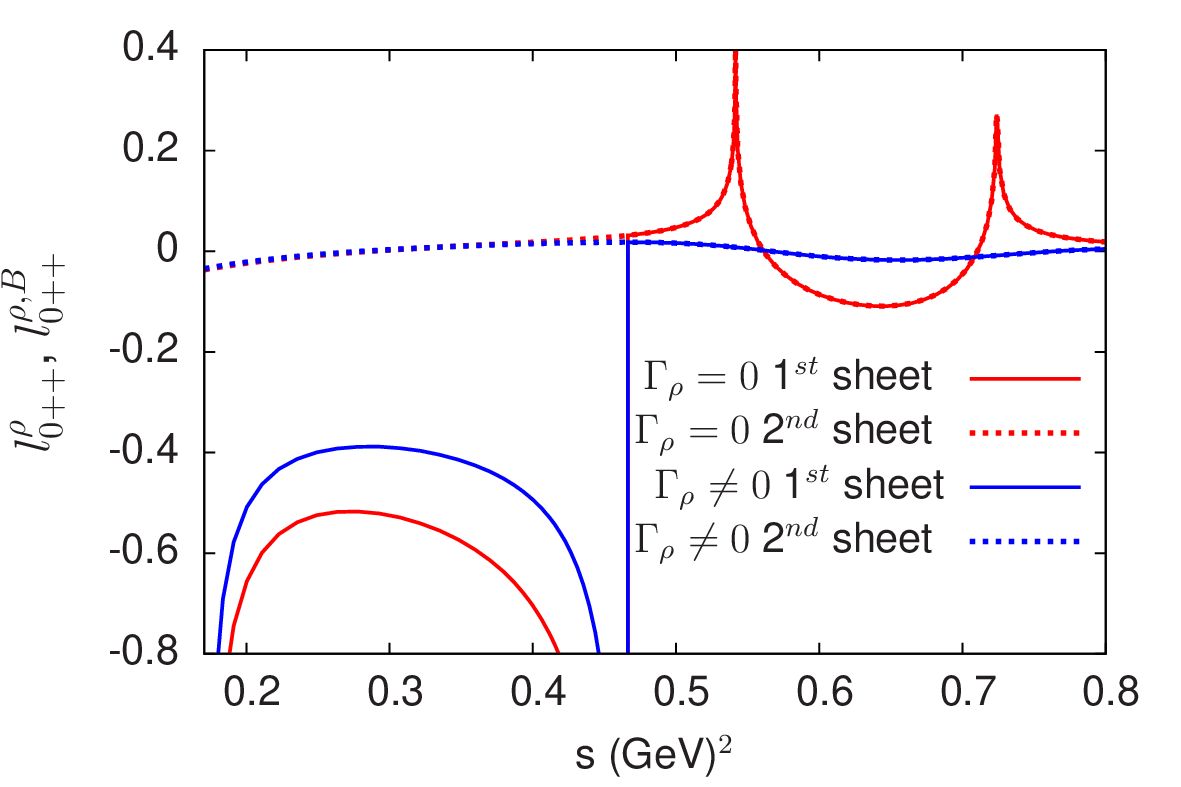}
\caption{\small $\rho$-meson exchange $J=0$ partial wave (real part) in the zero-width
  and finite width cases, showing the $B$ Riemann sheet extensions
  (dotted lines)
  in the region $s < m_+^2$.}
\lblfig{l0pp_rho}
\end{figure}
\section{Comparison with experiment}
Given a $T$-matrix satisfying two-channel unitarity and appropriate asymptotic
conditions the $\Omega$-matrix (and its inverse $D$) can be determined
numerically in a unique way. It is then only necessary to compute the left-cut
integrals which appear in eqs.~\rf{phi1phi2disp} and one obtains an expression
for the two $S$-wave $\phi$ decay amplitudes $l_{0++}$, $k^1_{0++}$ in terms of
two parameters $a_1$, $a_2$. An important check of the numerical
implementation (which we have performed) is to verify that the discontinuities
of the obtained solutions for $l_{0++}(s)$, $k^1_{0++}(s)$ across the unitarity
cut are exactly given by the unitarity equations~\rf{unitrelamplit}.

In the numerical results presented below we use the two-channel $T$-matrix
model introduced in ref.~\cite{Albaladejo:2015aca}. In this model, two-channel
unitarity is implemented using a $K$-matrix approach. The form of this
$K$-matrix is recalled in appendix~\sect{K-matrix}: it is 
constrained such that the low-energy expansion of the corresponding
$T$-matrix matches with the chiral expansion ($\chi$PT) up to NLO and
it involves six phenomenological parameters.  
Determinations of these parameters were obtained in
ref.~\cite{Lu:2020qeo} by performing fits to experimental measurements
of $\gamma\gamma\to \pi\eta$, $\gamma\gamma\to K_SK_S$ differential
cross sections in an energy range $E\lapprox 1.4$ GeV (a set of 688
data points was used).  These fits are based on Omn\`es
representations of the $\gamma\gamma$ $S$-wave amplitudes, exactly
analogous to eqs.~\rf{phi1phi2def},~\rf{phi1phi2disp} which depend on
two subtraction parameters in addition to the $T$-matrix
parameters. Two different sets of $T$-matrix parameters were actually
found in the $\gamma\gamma$ fits, giving approximately similar
$\chi^2$ minimum values. One of these sets gives rise to a ``narrow''
$a_0(1450)$ resonance and has some unphysical features, it will not be
considered here.

\subsection{Left-cut and right-cut integrals}
\begin{figure}[htb]
\centering
\includegraphics[width=0.499\linewidth]{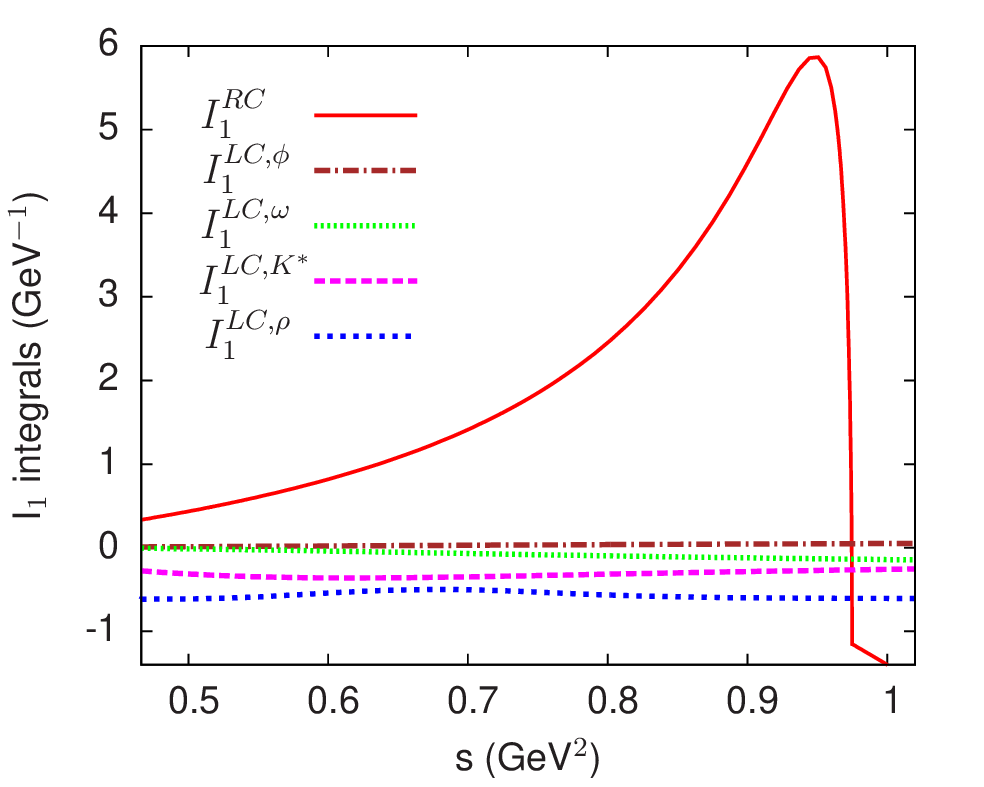}%
\includegraphics[width=0.499\linewidth]{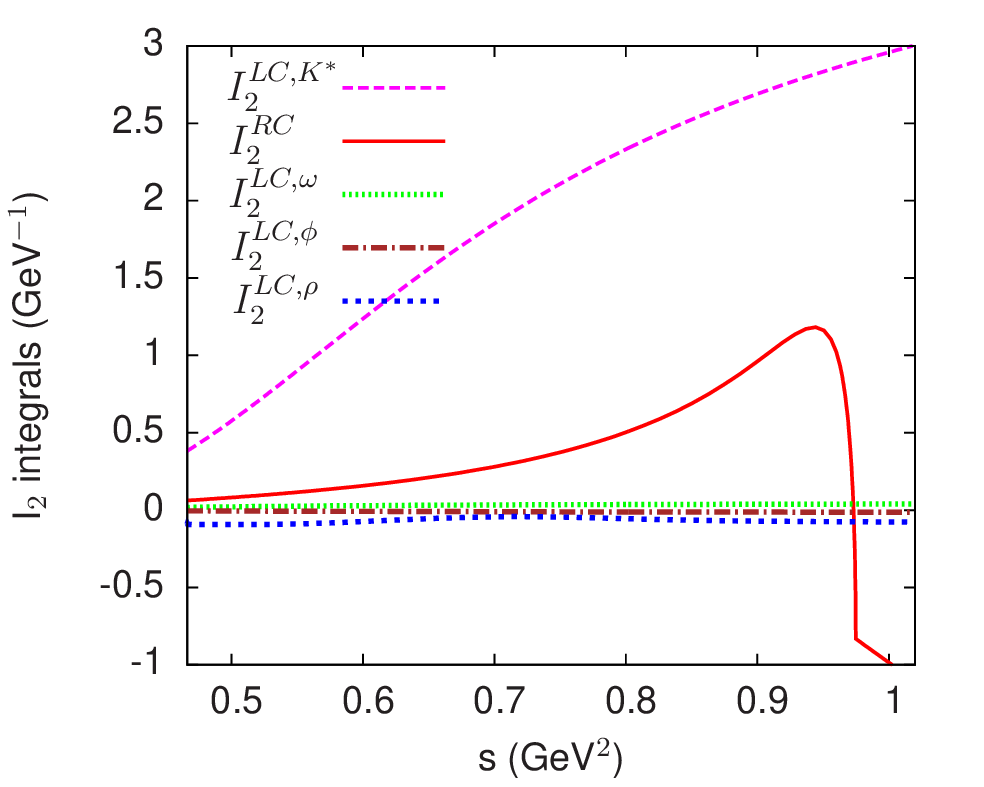}
\caption{\small Various contributions to the integrals $I_i^{LC}$, $I_i^{RC}$
(real parts) in the Omn\`es dispersive representation~\rf{omnesdisp}.}
\lblfig{ILC-IRC}
\end{figure}

Let us now re-write the expressions of the $S$-wave amplitudes in the
Omn\`es-dispersive approach
\be\lbl{omnesdisp}
\bp l_{0++}(s)\\
\bar{k}^1_{0++}(s) \\ \ep= (s-\qd)\bm{\Omega}(s)\bp
a_1 +I^{LC}_1(s,\qd)+ I^{RC}_1(s,\qd)\\
a_2 +I^{LC}_2(s,\qd)+ I^{RC}_2(s,\qd)\\ \ep
\en
in which we have defined $\bar{k}^1_{0++}(s)$ to be the part of the 
 the $I=1$ $K\Kbar$ amplitude
which vanishes at the soft photon point i.e.
\be\lbl{defkbar}
\bar{k}^1_{0++}(s)= k^1_{0++}(s,\qd)-
\alpha_B\Big[\frac{\beta(\qd)}{s-\qd} +\gamma(\qd)\Big]\ .
\en
The expressions of the four integrals $I^{LC}_i$, $I^{RC}_i$ were
given in eqs.~\rf{IRC_1}~\rf{IRC_2},~\rf{ILC_i},~\rf{ILC,V}. 
The integrals $I^{RC}_i(s)$ are shown in fig.~\fig{ILC-IRC}
along with the various vector meson contributions to the left-cut
integrals $I_i^{LC,V}$. The contributions from the $K^*$ are
enhanced by the fact that the effective coupling $\alpha_{K^*}$ is
one order of magnitude larger than the other $\alpha_V$ couplings
(see eq.~\rf{alpha_Vnumer}). The integrands of $I_i^{LC,V}$ are
proportional to the matrix elements $D_{i1}$, $D_{i2}$ of the $D$
matrix (eqs.~\rf{ILC,V},~\rf{ILC,rho0}).
At $z=0$ one has $D_{12}=D_{21}=0$, $D_{11}=D_{22}=1$ and in
the important integration region $|z| \lapprox 0.5$ GeV$^2$ the off
diagonal matrix elements $D_{12}, D_{21}$ are suppressed in
magnitude by roughly a factor of five as compared to the diagonal
ones $D_{11}, D_{22}$. This explains e.g. why $|I_2^{LC,K^*}| >>
|I_1^{LC,K^*}|$, $|I_2^{LC,K^*}| >>|I_2^{LC,\rho}|$.
A priori, one would expect the right-cut
integrals $I_i^{RC}$ to be dominant over the left-cut ones since they are
induced by the diverging part of the Born amplitude. This turns out to be
true in the case of the $I_1$ integrals  as can be seen from
fig.~\fig{ILC-IRC} (top). In the case of the $I_2$ integrals, however, one
sees from the figure that $I_2^{LC,K^*}$ is larger than the real part of
$I_2^{RC}$.

\subsection{Detailed comparison with KLOE results}\lblsec{datafits}
In the physical decay region, the $\phi\to\gamma\piz\eta$ helicity amplitudes
are dominated by the $S$-wave. The $J\ge1$ amplitudes are approximated by the
tree-level vector-exchange contributions. We have also made an estimate of the
$J=2$ amplitude induced by the $a_2(1320)$ tensor meson exchange in the
$s$-channel ($\phi\to\gamma a_2(1320)\to \gamma\pi\eta$, see
appendix~\sect{a2amplit}). This contribution turns out to be essentially
negligible in the physical region. 
Using these amplitudes, the  differential decay width of the $\phi\to
\gamma\pi^0\eta$ mode as a function of the $\pi\eta$ invariant mass
squared can be expressed as follows in terms of the
three independent helicity amplitudes
\be\lbl{diffGamma}
\ba{ll}
\dfrac{d^2\Gamma_{\phi\to\gamma\pi\eta}}
{d\sqrt{s}}& = 
\dfrac{(\qd- s) \lambda_{\pi\eta}(s) }{{384\pi^3} q^3 \sqrt{s}} 
{\displaystyle\int_{-1}^1}dz\,
\Big( \Big\vert l_{0++}(s) 
+{\displaystyle\sum_V}\tilde{L}^V_{++}(s,z)\Big\vert^2 \\[0.3cm]
\ &+\Big\vert{\displaystyle\sum_V} {L}^V_{+-}(s,z)\Big\vert^2 
+\Big\vert{\displaystyle\sum_V} {L}^V_{+0}(s,z)\Big\vert^2
\Big)
\ea\en
where $L^V_{\lambda\lambda'}$ are tree-level vector-exchange amplitudes
with $\tilde{L}^V_{++}(s,z)\equiv {L}^V_{++}(s,z)-l^V_{0++}(s)$. 
We will compare the results from this theoretical model with the
experimental measurements of the single differential decay width
performed by the KLOE collaboration~\cite{Ambrosino:2009py}. 
They are presented in table 5 of ref.~\cite{Ambrosino:2009py} in the
form of 49 data points which have been corrected from the background
and the energy resolution effects. We will not include the last two
points in the $\chi^2$ (see the discussion below).
The $\chi^2$ is then computed from the following formula,
\be
\chi^2(\phi)=\sum_{i=1}^{47}  \frac{1}{\sigma_i^2} 
\left( \frac{1}{\Gamma_\phi}\frac{d\Gamma_{exp}}{dE_i}
-\frac{1}{\Gamma_\phi}\int_{E_i-\Delta{E}/2}^{E_i+\Delta{E}/2} d\sqrt{s'}
\frac{d\Gamma_{th}}{d\sqrt{s'}} \right)^2
\en
where $\sigma_i$ is the error on the $i^{th}$ data point,
$\Delta{E}=6.35$ MeV is the size of the energy bins, $\Gamma_\phi$ is
the total width of the $\phi$ (we take $\Gamma_\phi=4.26$ MeV and the
value of the photon virtuality $q=1020$ MeV as in
ref.~\cite{Ambrosino:2009py}).

\begin{figure}[hbt]
\centering
\includegraphics[width=0.60\linewidth]{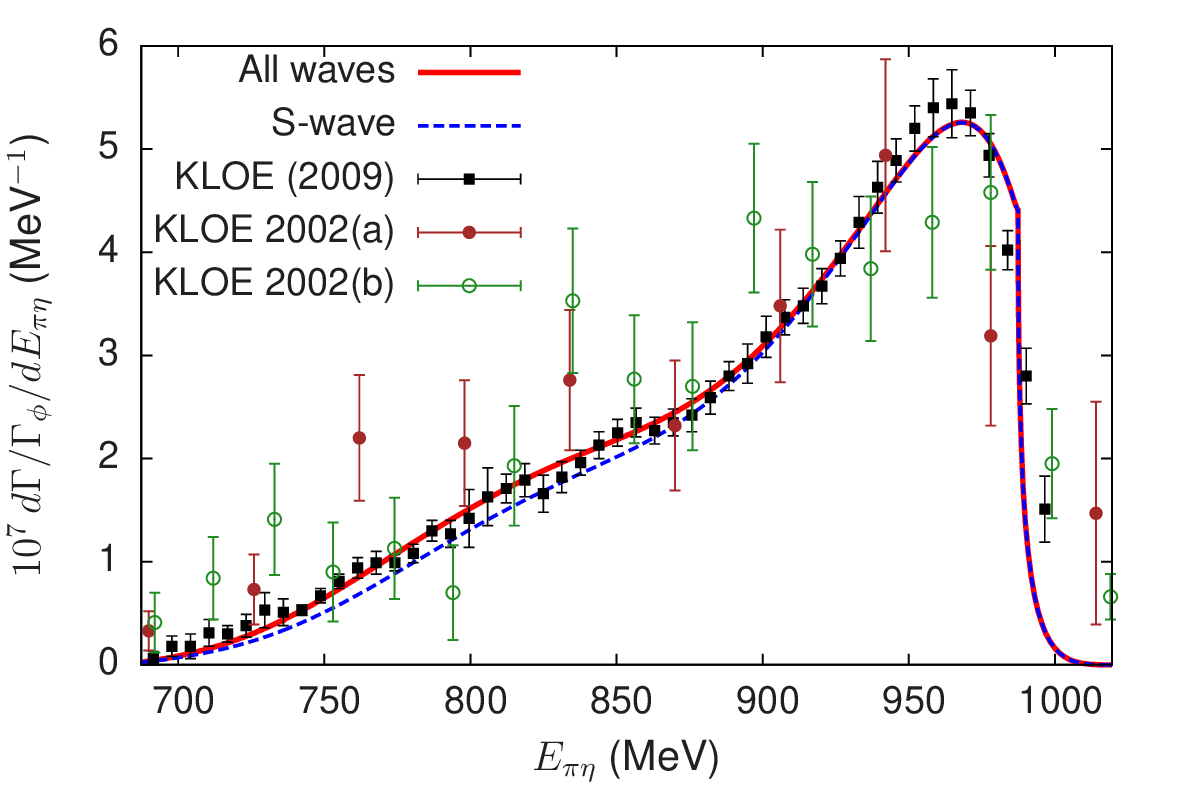}
\caption{\small Normalised differential energy distribution of the $\phi\to
  \gamma\piz\eta$ decay width. The dashed blue curve is the result
from the dispersive $S$-wave amplitude and the solid red curve is
the result including the $J\ge 1$ partial waves. The experimental points
are results from the KLOE collaboration: black squares are
from~\cite{Ambrosino:2009py}, 
brown filled circles and green empty circles are from
ref.~\cite{Aloisio:2002bsa} (the $\eta$ being detected via the $\eta\to3\pi$ 
and the $\eta\to2\gamma$ modes respectively).}   
\lblfig{diffwidth}
\end{figure}
The determination of the $T$-matrix parameters performed in
ref.~\cite{Lu:2020qeo} is based on a dispersive Omn\`es description of the
$\gamma\gamma\to \pi\eta, K\Kbar$ amplitudes involving two
subtractions parameters. One of these was fixed by assuming a given
position of the Adler zero  $s_A^{\gamma\gamma}$ in the
$\gamma\gamma\to \pi\eta$ $S$-wave amplitude. Using the $T$-matrix
which corresponds to $s_A^{\gamma\gamma}=\metad$ in the computation of
the $\phi$ decay amplitudes and fitting the two parameters $a_1$,
$a_2$ to the KLOE data on gets $\chi^2(\phi)/N_{dof}=58/45$. 
Taking a slightly larger value for the $\gamma\gamma$ amplitude Adler
zero:\footnote{ This value gives a somewhat better description of the
  $\gamma\gamma$ data: $\chi^2(\gamma\gamma)=422$ using 688 data points
  (instead of $\chi^2(\gamma\gamma)=439 $) and also a better description of
  the decay $\eta\to \piz\gamma\gamma$ width distribution (see fig.10 in
  ref.~\cite{Lu:2020qeo} )} $s_A^{\gamma\gamma}=\metad+3\mpid$ gives a
$T$-matrix which provides an even better fit to the $\phi$ decay data
with $\chi^2(\phi)/N_{dof}=40/45$ and the values of $a_1$, $a_2$ are
\be\lbl{a1a2values}
a_1=0.24\pm 0.01,\quad
a_2=-1.74\pm 0.03\quad (\hbox{GeV}^{-1}) .
\en
The ability to reproduce the $\phi$ decay data in the whole physical
energy region with only two fit parameters\footnote{For comparison,
  the fits performed in ref.~\cite{Ambrosino:2009py} used either 5
  parameters (KL model~\cite{Achasov:1987ts,Achasov:2002ir}) or 7
  parameters (NS model~\cite{Isidori:2006we})} indicates a good
compatibility between these data and the $\gamma\gamma$ scattering
data concerning the final-state rescattering $T$-matrix.

\begin{figure}[ht]
\centering
\includegraphics[width=0.49\linewidth]{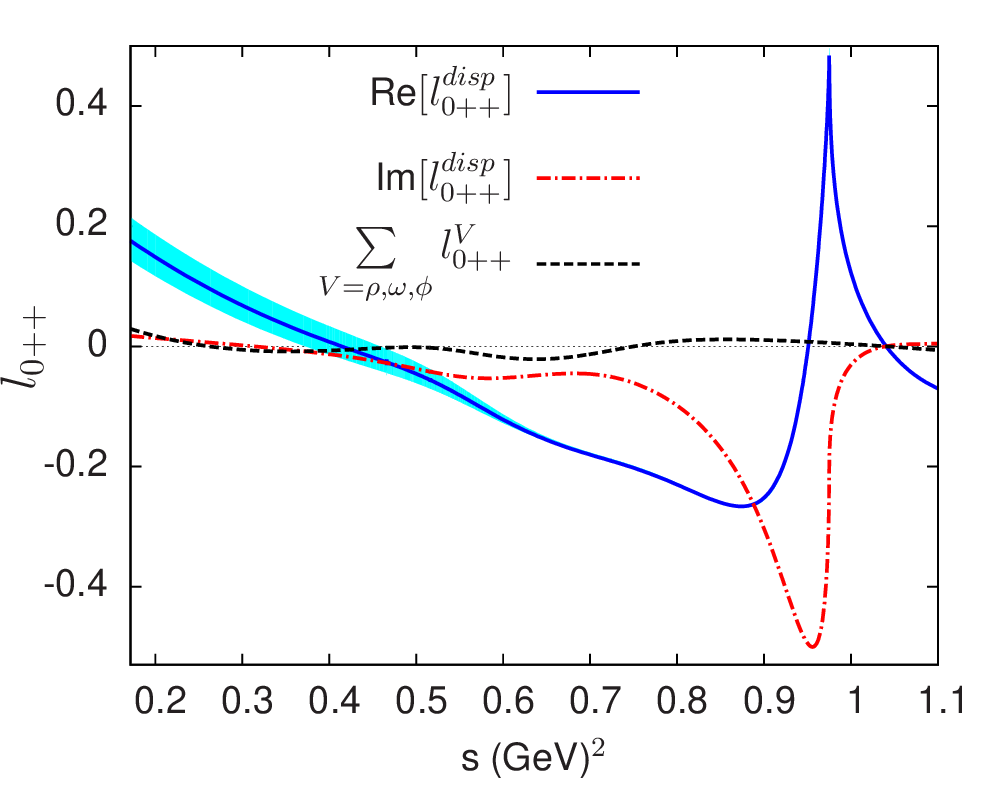}%
\includegraphics[width=0.49\linewidth]{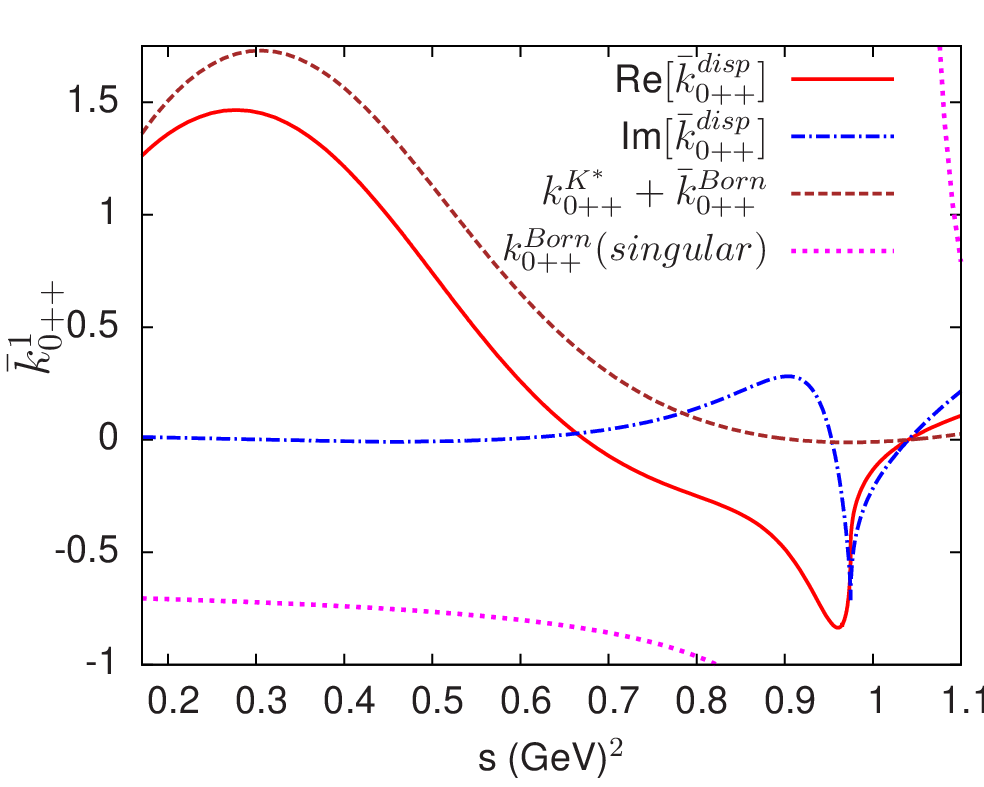}
\caption{\small The figure on the left displays the result for the dispersively
constructed $l_{0++}$ amplitude. Below the $\pi\eta$ threshold the
extension to the $B$-Riemann sheet is shown (see sec~\sect{B-Riemann}).
The figure on the right displays the result for the $\bar{k}^1_{0++}$
amplitude. Also shown for comparison
are the tree-level vector-exchange amplitudes and the singular part of the
Born amplitude. The cyan band shows the variation of $l_{0++}$ when the
coupling constants $C_{VP\gamma}$ and $g_{\phi{VP}}$ are varied.}
\lblfig{l0ppk0pp}
\end{figure}
%
The differential decay width corresponding to this fit is displayed in
fig.~\fig{diffwidth} showing separately the contribution of the $S$-wave.  One
sees a small but visible contribution from the higher waves in the energy
region below 900 MeV. This contribution is dominated by the $J=1$ amplitude
which is allowed for $\gamma^*\to \gamma\pi\eta$ when $\qd\ne 0$.
The two $S$-wave amplitudes $l_{0++}$ and $\bar{k}^1_{0++}$ obtained from this
fit are plotted in fig.~\fig{l0ppk0pp} and compared with the tree-level
vector-exchange amplitudes. The amplitude $\bar{k}^1_{0++}$ is seen to
qualitatively agree in size with the sum of the $K^*$ exchange and the regular
part of the Born amplitude for energies below 0.8 $\hbox{GeV}^2$.  In
contrast, the behaviour of the $\pi\eta$ amplitude $l_{0++}$ is dominated by
the rescattering mechanism even at low energy. In the energy region below the
threshold $s \le m_+^2$ the $B$ Riemann sheet extension $l_{0++}^{(B)}$ is
shown (see sec.~\sect{B-Riemann}), which is expected to display an Adler zero
$s_A^\phi=\metad+O(\mpid)$. The figure shows that this is indeed the case. The
central value of this zero is located at $s_A^\phi=\meta^2+6.2\mpid $ and its
position varies in the range $[\meta^2+3.7\mpid, \meta^2+8.1\mpid]$ when
varying the input values of the coupling constants which control the left-hand
cut discontinuities. These features indicate that the fitted values of the two
parameters $a_1$, $a_2$ have a physically reasonable size.
%

\begin{figure}[hbt]
\centering
\includegraphics[width=0.55\linewidth]{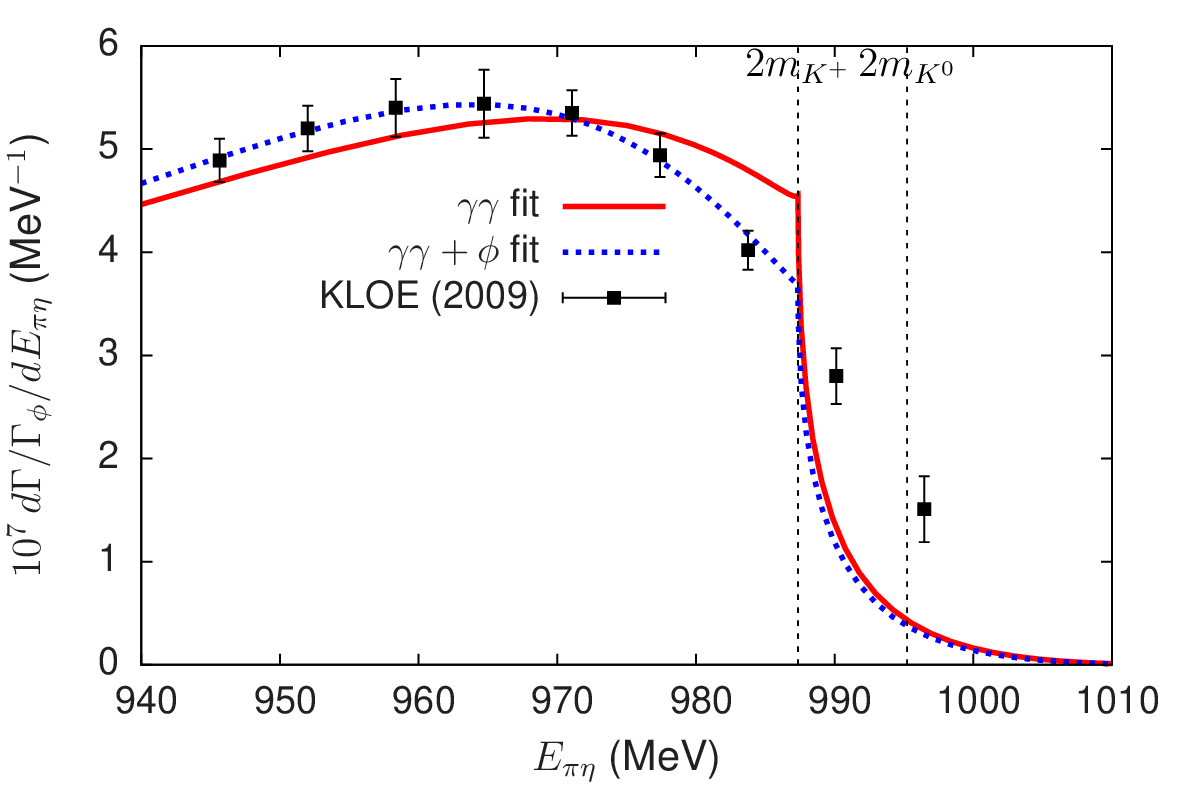}
\caption{\small Results for $d\Gamma_{\phi\to\gamma\pi\eta}/dE_{\pi\eta}$ when the
$T$ matrix is determined from $\gamma\gamma$ data only (solid red curve) and
  when it is determined by combining the $\gamma\gamma$ and the $\phi$ data
  (dotted blue line).}
\lblfig{diffwidth_end}
\end{figure}
\subsection{Combined $\gamma\gamma$ and $\phi$ decay fits}
We have seen that the $T$-matrix determined from $\gamma\gamma$ data
is compatible with the $\phi$ decay data from KLOE. It is interesting
to study how the $T$-matrix parameters would be modified if the two
data sets were combined. We have performed such a fit in which we have
increased the weight of the $\phi$ decay data by a factor of two. 
In this way, the following total $\chi^2$ values are
obtained,
\be
\chi^2[\phi]=15.6,\quad \chi^2[\gamma\gamma]=445.8\quad (\hbox{combined fit})
\en
showing a significant improvement in the $\phi$ data $\chi^2$. At the
same time the $\chi^2$ value of the $\gamma\gamma$ data remains acceptable.
For comparison, the fit including only $\gamma\gamma$ data (with
$s_A^{\gamma\gamma}=\metad+3\mpid$) gives 
\be
\chi^2[\phi]=39.8,\quad \chi^2[\gamma\gamma]=421.8\quad 
(\gamma\gamma\ \hbox{fit})\ .
\en
The $\phi$ differential decay width in the two fits 
differ essentially in the higher energy region $E_{\pi\eta} \gapprox 940$
MeV. This is illustrated in fig.~\fig{diffwidth_end} which also shows
the location of the $\Kp\Km$, $\Kz\Kzb$ thresholds.  One observes, in
particular, a significant improvement on the last two points below
the $\Kp\Km$ threshold (points 46 and 47) in the combined
fit. However, the two points which are above (points 48 and 49) are
not well described in either one of the fits. One observes that the
energy of the point 48 is located in between the $\Kp\Km$ and the
$\Kz\Kzb$ thresholds. It is possible that the discrepancy could be
explained by isospin breaking effects which were predicted to be
enhanced in this energy region~\cite{Achasov:1979xc}.  In our model,
isospin symmetry is assumed, and we have taken $m_K=m_{K^+}$ for 
the kaon mass in order to have the correct mass in the Born amplitude. Because
of the importance of the Born amplitude in the rescattering dynamics it seems
plausible that the cusp at the $K^+K^-$ threshold should be significantly more
pronounced than the one at the $\Kz\Kzb$ threshold. A complete treatment of
isospin breaking effects is difficult since they induce couplings between the
$I=1$ channels considered here and $I=0$ as well as $I=2$ channels. One must
then use a six-channel $T$-matrix and it is clearly not possible to determine
all the $T$-matrix elements in a model independent way. An example of a
plausible modelling has been proposed e.g. in ref.~\cite{Hanhart:2007bd}.

On the experimental side, a clear observation of a two-particle
threshold effect requires a very good energy resolution (for an
example, see ref.~\cite{Batley:2005ax}). The data points tabulated in
ref.~\cite{Ambrosino:2009py} are obtained after implementing a
procedure for unfolding the energy resolution effects and subtracting
the background. It is possible that this procedure becomes somewhat
inaccurate near the tip of the Dalitz plot region.

Details of the numerical values of the $T$-matrix parameters
corresponding to different fits are given in
appendix~\sect{K-matrix}. Fig.~\fig{phaseshifts} shows the phase
shifts and inelasticity parameter of the $S$-matrix corresponding to
two different fits. We observe that in these determinations the phase
shift $\delta_{\pi\eta}$ increases above the $K\Kbar$ threshold while
$\delta_{K\Kbar}$ decreases. This pattern differs from the one found
in the lattice QCD simulation~\cite{Dudek:2016cru}.
\begin{figure}
\centering
\includegraphics[width=0.60\linewidth]{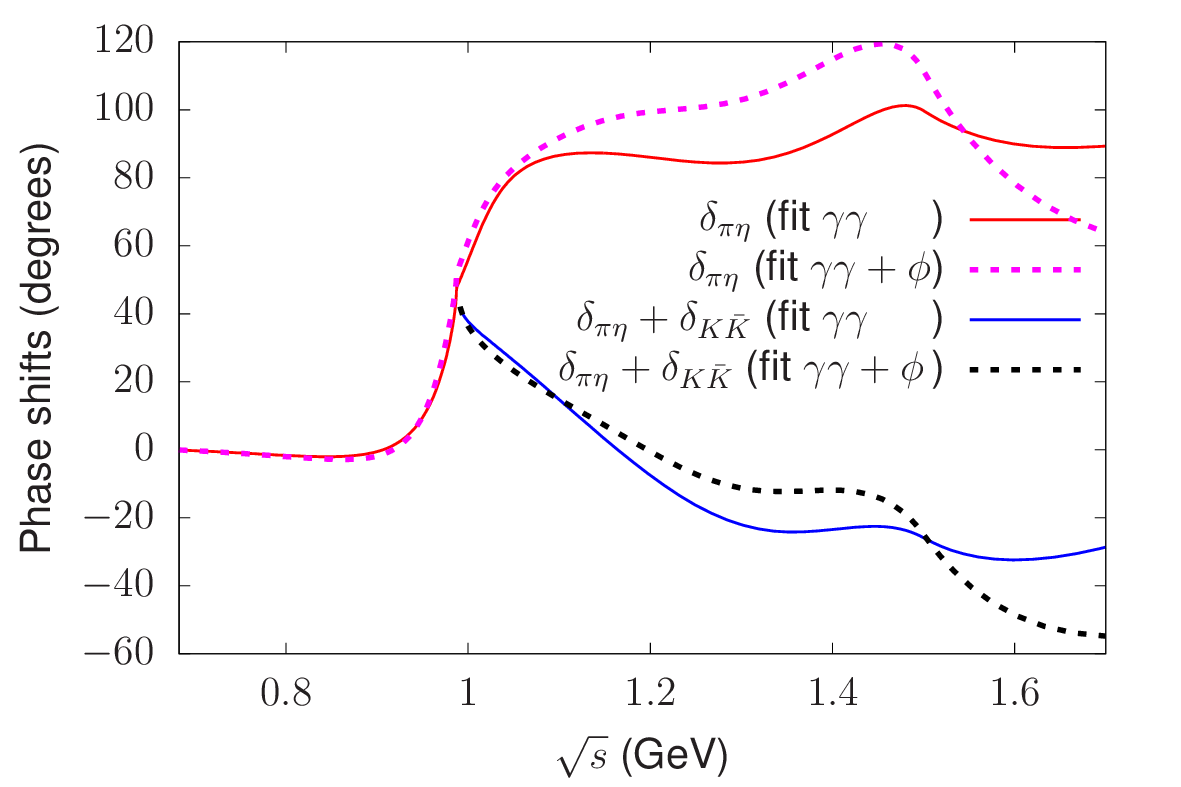}
\includegraphics[width=0.60\linewidth]{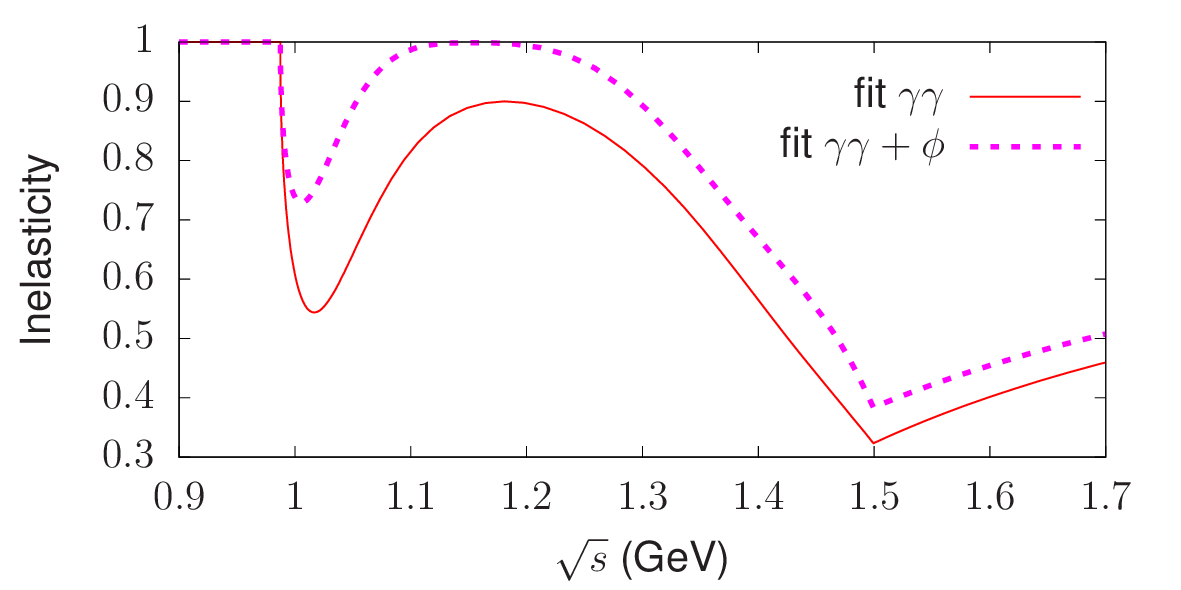}
\caption{\small  Comparison of results from two different fits
  corresponding to the parameters shown in the last two columns of
  table~\Table{paramvalues}. The upper figure shows the $\pi\eta$
  phase shift and the sum of the $\pi\eta$ and $K\Kbar$ phase shifts
  and the lower figure shows the inelasticity parameter.}
\lblfig{phaseshifts}
\end{figure}

\subsection{$a_0(980)$ complex pole and couplings}\lblsec{a0results}
In the determinations of the $\pi\eta-K\Kbar$ two-channel $T$-matrix
performed in ref.~\cite{Lu:2020qeo} based on analysing $\gamma\gamma$
scattering data, the $a_0(980)$ resonance was always found to
correspond to a pole of the $T$-matrix located on the second Riemann
sheet. No pole was found on the third or fourth Riemann sheet with a
real part close to 1 $\hbox{GeV}^2$.
We recall the formulae which define the second sheet extension of the
$T$-matrix elements (see e.g.~\cite{Barton:1965})
\be
T_{11}^{II}(z)=\frac{T_{11}(z)}{S_{11}(z)},\quad
T_{12}^{II}(z)=\frac{T_{12}(z)}{S_{11}(z)},\quad
T_{22}^{II}(z)=T_{22}(z)+\frac{2\tilde{\sigma}_{\pi\eta}(z)
\left(T_{12}(z)\right)^2} {S_{11}(z)}
\en
with
\be\lbl{S_11def}
S_{11}(z)=1-2\tilde{\sigma}_{\pi\eta}(z)T_{11}(z),\quad
\tilde{\sigma}_{\pi\eta}(z)=\sqrt{ (m^2_+-z)(z-m^2_-)}/z\ .
\en
It is indeed easy to verify that the matrix elements $T_{ij}^{II}(s)$
satisfy the continuity equation across the cut
\be
T_{ij}^{II}(s-i\epsilon)= T_{ij}(s+i\epsilon),\quad
(\meta+\mpi)^2 \le s \le 4\mkd\ .
\en
Simple definitions of the $a_0$ coupling constants
can be associated with residues of the $T$-matrix poles,
\be
\left.16\pi T_{11}^{II}(z)\right\vert_{z\to s_{a_0}}=
\frac{g^2_{a_0\pi\eta}}{s_{a_0}-z},\quad
\left.16\pi T_{12}^{II}(s)\right\vert_{z\to s_{a_0}}=
\frac{g_{a_0\pi\eta}\, g_{a_0K\Kbar}}{s_{a_0}-z}\ .
\en
A formal definition of the width $\Gamma_{a_0\to \pi\eta}$ can be
associated with the coupling $g_{a_0\pi\eta}$,
\be
\Gamma_{a_0\to \pi\eta}=\frac{\vert g_{a_0\pi\eta}   \vert^2}{16\pi
  m^3_{a_0}}\lambda_{\pi\eta}(m^2_{a_0})\ .
\en
The $a_0$ coupling to $K^+K^-$, which is often quoted, 
is related to $g_{a_0K\Kbar}$ by
\be
g_{a_0K^+K^-}=-\frac{g_{a_0K\Kbar}}{\sqrt2}\ .
\en
The magnitude of the couplings  $g_{a_0\pi\eta}$, $g_{a_0K^+K^-}$ as defined
above should be approximately equal to those defined in
refs.~\cite{Achasov:1987ts,Isidori:2006we}.

The analytic extensions of the $\phi$ radiative decay $S$-wave amplitudes
to the 2nd Riemann sheet are defined as
\be
l_{0++}^{II}(z)= \frac{l_{0++}(z)}{S_{11}(z)},\quad
k_{0++}^{II}(z)= k_{0++}(z) 
+\frac{2\tilde{\sigma}(z)T_{12}(z)l_{0++}(z)}{S_{11}(z)}\ 
\en
which shows that they should display an $a_0$ pole.
A  coupling constant $g_{\phi a_0\gamma}$ can be defined from the residue
\be
\left. l_{0++}^{II}(z)\right\vert_{z\to s_{a_0}}=\frac{1}{2}e \frac{ (\qd-z)
g_{\phi a_0\gamma}   g_{a_0\pi\eta}}{s_{a_0}- z}\ .
\en
This definition (which takes into account the presence of the nearby soft
photon zero) matches to the one introduced in ref.~\cite{Isidori:2006we}.

The central values of the $a_0(980)$ mass, width and couplings, corresponding
to several fits are collected in table~\Table{a0reson}. The last column
corresponds to a fit which combines the $\gamma\gamma$ data and the $\phi$
data. In this fit, the $a_0$ mass is somewhat smaller as well as the coupling
$g_{\phi a_0\gamma}$. As discussed in~\cite{Lu:2020qeo} an important source of
uncertainty is generated by the ``fixed'' parameters in the $T$-matrix, mainly
the couplings $L_i$ and the ratio $c_m/c_d$ (see eqs.~\rf{K6}~\rf{g1g2}), as
well as the value of the Adler zero in the $\gamma\gamma\to \pi\eta$
amplitude. We have performed fits in which these parameters are varied around
their central values in the same way as in~\cite{Lu:2020qeo} except for
$s_A^{\gamma\gamma}$ which we vary here in the range
$[\meta^2,\meta^2+6\mpid]$. As a result of this, we would quote the following
determination of the mass/width from these data
\be
\ba{ll}
m_{a_0}      & = 989^{+12}_{-3}\ (\MeV)\\[0.2cm]
\Gamma_{a_0} & =  76^{+14}_{-6}\ (\MeV)\\
\ea\en
and for the $a_0$ couplings
\be
\ba{ll}
g_{a_0\eta\pi}= & 2.2\pm0.2\ (\hbox{GeV})\\[0.1cm]
g_{a_0\Kp\Km} = & 2.8\pm0.2\ (\hbox{GeV})\\[0.1cm]
g_{\phi a_0\eta\pi}= & 2.3 ^{+0.7}_{-0.3}\ (\hbox{GeV}^{-1})\ .
\ea
\en

\begin{table}
\centering \bt{c|c|c|c}\hline\hline 
\  & $\gamma\gamma$ fit &  $\gamma\gamma$ fit   & $\gamma\gamma+ \phi$ fit\\
\  & $(s^{\gamma\gamma}_A=\metad)$     &  $(s^{\gamma\gamma}_A=\metad+3\mpid)$
&  $(s^{\gamma\gamma}_A=\metad+3\mpid)$    \\ \hline
\TT   $m_{a_0}$ (MeV)    &  1000.6  &   1002.8   &    988.8   \\ 
$\Gamma_{a_0}$ (MeV)     &    71.1  &     79.1   &     75.6   \\ 
$g_{a_0\eta\pi}$ (GeV)     &     2.14  &     2.35  &     2.07   \\ 
$g_{a_0\Kp\Km}$ (GeV)     &      2.85  &     2.89  &    2.74    \\ 
$g_{\phi a_0\gamma}$  (GeV$^{-1}$)&  3.26   &       3.21 &   1.96  \\  \hline\hline 
\et 
\caption{\small Mass and width of the $a_0(980)$ resonance associated with the
complex pole position ($\sqrt{s_{a_0}}= m_{a_0}-i\Gamma_{a_0}/2$) and
the related coupling constants in several fits.  }
\lbltab{a0reson} 
\end{table}

\begin{figure}[hbt]
\centering
\includegraphics[width=0.55\linewidth]{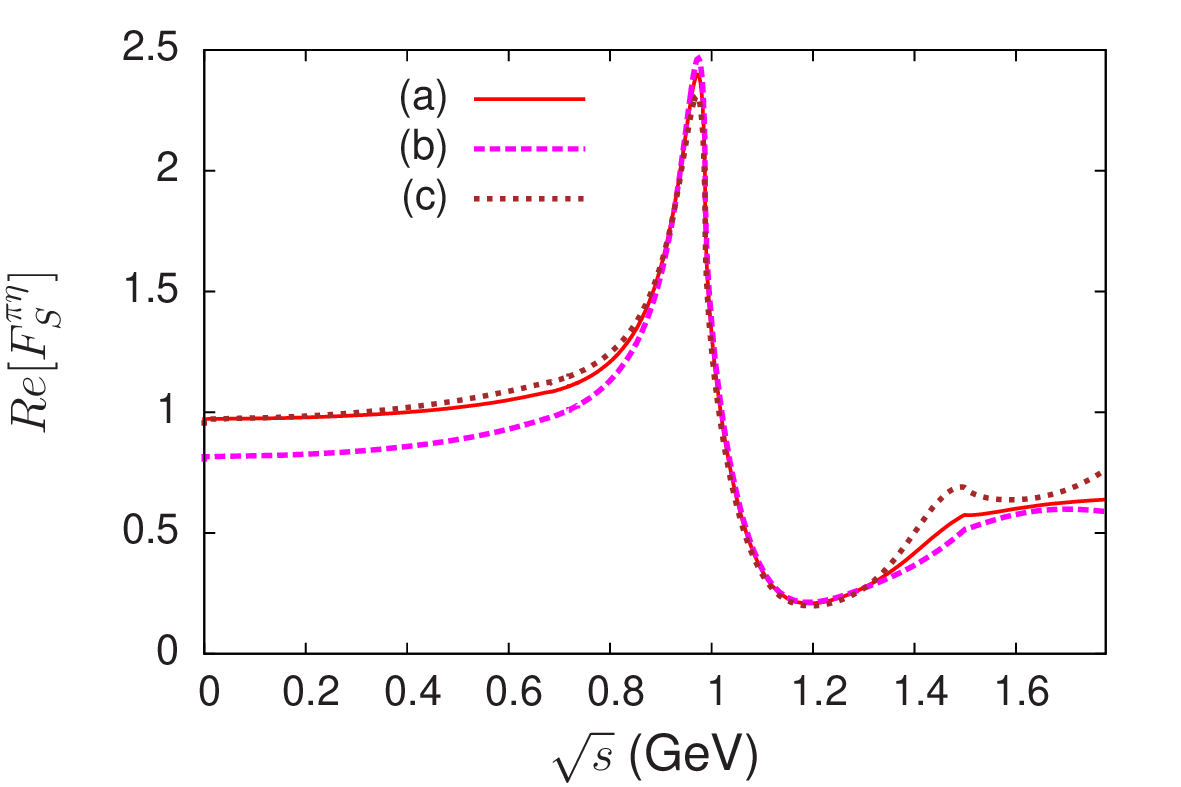}
\includegraphics[width=0.55\linewidth]{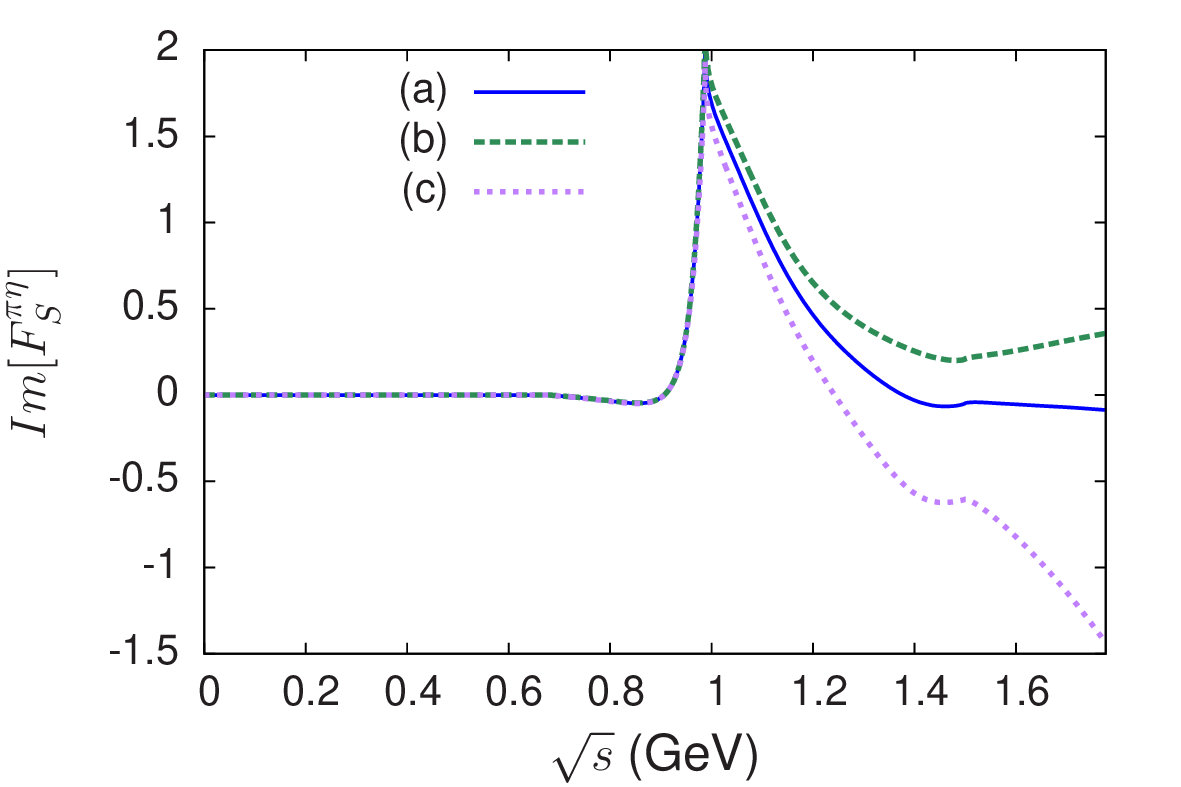}
\caption{\small Dispersive results for the real part (upper plot) and the
  imaginary part (lower plot) of the scalar form factor $F_S^{\pi\eta}$. The
  labelling is as follows: (a) minimal model with $O(p^2)$ values at $s=0$ (b)
  minimal model with $O(p^4)$ values at $s=0$ and (c) non-minimal model
  enforcing $O(p^4)$ values for both $F_S^{PQ}(0)$ and the derivatives
  $\dot{F}_S^{PQ}(0)$.} 
\lblfig{FFactors}
\end{figure}
\subsection{Scalar $I=1$ form factors}\lblsec{scalarff}
As already mentioned, the $I=1$ $\pi\eta$ and $K\Kbar$ scalar form
factors are linearly related to the Omn\`es matrix elements which have
been probed by the $\gamma\gamma\to \pi\eta$ and the
$\phi\to\gamma\pi\eta$ processes.
In view of their applications to $\tau$ decays, it is convenient to define the
scalar form factors from the matrix elements of the charged scalar current
$\bar{u}d(x)$ as
\be
\ba{l}
B_0\,F_S^{\pi\eta}(s)\equiv 
  \braque{\pip(p_\pi)\eta(p_\eta)\vert \bar{u}d(0)\vert 0}\\[0.2cm] 
B_0\,F_S^{K\bar{K}}(s)\equiv 
  \braque{\Kp(p_\Kp)\Kzb(p_\Kz)\vert \bar{u}d(0)\vert 0}
\ea
\en
where $s= (p_\pi+p_\eta)^2=(p_\Kp+p_\Kz)^2$ and $B_0$ is the $O(p^2)$ chiral
coupling proportional to the quark condensate~\cite{Gasser:1984gg}. In a
minimal construction, the form factors are related to the Omn\`es matrix
simply by
\be\lbl{FFfromomnes}
\bp
F_S^{\pi\eta}(s)\\[0.2cm]
F_S^{K\bar{K}}(s)\\
\ep= \bp 
\Omega_{11}(s) & -\Omega_{12}(s)\\[0.2cm]
-\Omega_{21}(s) & \Omega_{22}(s)\\
\ep \bp
F_S^{\pi\eta}(0)\\[0.2cm]
F_S^{K\bar{K}}(0)\\
\ep
\en
(the minus sign being caused by $\ket{\pip\eta}=-\ket{I=1,I_z=1}$). 
At leading
chiral order, the values at $s=0$ are given by 
\be
F_S^{\pi\eta}(0)=\sqrt{\frac{2}{3}},\quad 
F_S^{K\Kbar}(0)=1\quad  (O(p^2))\ .
\en
Expressions for the NLO corrections can be found
e.g. in~\cite{Albaladejo:2015aca,Shi:2020rkz}. Using the determination 
BE14~\cite{Bijnens:2014lea} for the chiral coupling constants $L_i$ (as
used also here for the $T$-matrix) one finds for the central values
\be
F_S^{\pi\eta}(0)=0.9725\ ,\quad
F_S^{K\Kbar}(0)=0.8440\quad  (O(p^4))  \ . 
\en
Other chiral constraints that one might consider concern the scalar radii
\be
\braque{r^2}_S^{PQ}\equiv \dot{F}_S^{PQ}(0)/6{F}_S^{PQ}(0)
\en
which are not difficult to compute at order $p^4$
(see~\cite{Albaladejo:2015aca}). Using the BE14 value for the coupling
$L_5$, $L^r_5=1.01\cdot10^{-3}$, one gets
\be\lbl{scalrad}
\braque{r^2}_S^{\pi\eta}=0.069\ \hbox{fm}^2,\quad
\braque{r^2}_S^{K\Kbar} =0.112\ \hbox{fm}^2\quad  (O(p^4))\ .
\en
The minimal dispersive model gives a result which is too small by 50\%
for the $\pi\eta$ radius and too large by 40\% for the $K\Kbar$ one as
compared to the $O(p^4)$ values. While these differences can be
ascribed, in part, to higher order corrections, it seems interesting
to consider a non-minimal dispersive model in which one enforces
exactly the $O(p^4)$ values for both $F_S^{PQ}(0)$ and
$\dot{F}_S^{PQ}(0)$. This is easily achieved by replacing
${F}_S^{PQ}(0)$ in eq.~\rf{FFfromomnes} by linear functions of
$s$, $F_S^{PQ}(0)(1+\lambda^{PQ}\,s)$, and adjusting the slope
parameters $\lambda^{PQ}$.
Fig.~\fig{FFactors} shows results for the real and imaginary parts of the
$\pi\eta$ form factor in the minimal dispersive model using either LO or NLO
values for $F_S^{PQ}(0)$ and the result from the non-minimal
model. Differences between the models remain rather small in the region of the
$a_0(980)$ resonance which is important for the $\tau$ decay. 
Our results are
rather similar to those presented recently in ref.~\cite{Shi:2020rkz} but
differ somewhat from ref.~\cite{Escribano:2016ntp}.

\section{Summary and conclusions}
We have re-examined the $\phi$ radiative decay amplitudes $\phi\to
\gamma\pi\eta, \gamma K\Kbar$ based on coupled-channel Omn\`es-type dispersive
representations for the $S$-waves, which seems not to have been done
previously.  By construction, these amplitudes have the correct analytic
structure, with a left-hand cut having several components in addition to the
unitarity cut, and the representation involves explicit integrations along
these cuts.
The discontinuities across the left-hand cut components were assumed to be
dominated by the contributions from the light vector-meson resonance
exchanges.  The coupling constants which are needed can be determined
approximately (including the relative signs) by combining experimental inputs
with flavour and chiral symmetry arguments. This approach, which uses only the
discontinuities of the vector-exchange amplitudes has the advantage of being
insensitive to the off-shell behaviour of the resonance propagators
(polynomial ambiguities, form factors) and allows to take into account the
effects of the resonance widths in a numerically fast way.  

The kaon exchange  Born contribution to the $\phi\to\gamma\Kp\Km$ amplitude
plays an important role in the dynamics, as is well known, because it carries
a pole when $s= m^2_\phi$. It proved convenient to split this
amplitude into a singular part and a regular part (which vanishes in the soft
photon limit). The integrals involving the first part can be computed
analytically in terms of matrix elements of the $D$ matrix. In practice, we
describe the $\pi\eta/K\Kbar$ scattering using a two-channel unitary 
$T$-matrix model proposed in~\cite{Albaladejo:2015aca}, which has both left and right-hand cuts,
and involves six phenomenological parameters. The corresponding $\Omega$ (and
$D$) matrices are computed numerically by solving a set of Muskhelishvili
integral equations. In some previous work an apparently different
representation was used (e.g.~\cite{Achasov:1987ts,Nussinov:1989gs,LucioMartinez:1990uw,Close:1992ay,Bramon:1992ki,Oller:1998ia,Marco:1999df} ) which involves a one-loop triangle
function and the $T$-matrix itself. As shown in appendix~\sect{kaonloopderiv}  
this representation holds only for restricted classes of
$T$-matrices which, in particular, have no left-hand cuts.

We have shown that a rather good description of the
experimental data on $\phi\to\gamma\pi\eta$ can be obtained fitting only two
subtraction parameters, while using 
in the $T$-matrix a set of parameters determined
previously from photon-photon scattering data. The 
fitted values of the two subtraction parameters were shown to be
physically reasonable, such that the $\phi\to \gamma\pi\eta$ amplitude
displays an Adler zero and the $\phi\to \gamma K\Kbar$ amplitude is
dominated by the sum of the $K^*$-exchange and the regular part of the
$K$-exchange amplitudes at low energy.

These results seem to confirm the validity of the two-channel $T$-matrix model
and the corresponding $\Omega$ matrix. As an application, we give results for
the $\pi\eta$ and the $(K\Kbar)_{I=1}$ scalar form factors (which are simply
related to the $\Omega$ matrix), they were shown to
be well determined in the region of the $a_0(980)$ resonance using $O(p^4)$
chiral constraints at $s=0$.  The $\pi\eta$ form factor is measurable, in
principle, from the $\tau\to \pi\eta\nu$ decay mode. Finally, we give some
results on the $a_0(980)$ complex pole, performing combined fits of
$\gamma\gamma$ data and $\phi$ radiative decay data.

\bigskip
\noindent{\Large\bf Acknowledgements:}\\
This work is supported  by the European Union's Horizon2020 research
and innovation programme (HADRON-2020) under the Grant Agreement n$^\circ$
824093.

\appendix

\section{Relation between the Omn\`es and the kaon-loop
  representations}\lblsec{kaonloopderiv}
In this appendix we show that the Omn\`es representations for the
$\gamma\gamma^*\to \pi\eta$ amplitudes can be recast, for certain
classes of $T$-matrices, in a form which involves  one-loop
triangle functions. The kaon loop function, in particular, can be
written in dispersive form as follows\footnote{Analytical expressions can be
  found in refs.~\cite{Achasov:1987ts,Close:1992ay}}
\be
I_K^{loop}(s,\qd)=\frac{s-\qd}{\pi}\int_{4\mkpd}^\infty ds'\,
\frac{\sigma_{\Kp}(s')\, k^{1,B}_{0++}(s')}{(s'-s)(s'-\qd)}\ .
\en
Let us illustrate this in the case of the 
$T$-matrices based on unitarised $\chi$PT used in
refs.~\cite{Oller:1998ia,Marco:1999df}.  In
these models the $T$-matrices have no left-hand cut and the $D$
and $\Omega$ matrices can be expressed explicitly as follows
\be\lbl{Dmatrixoset}
\bm{D}(s)\equiv \bm{\Omega}^{-1}(s)=
\bm{1}- \bm{T}^{(2)}(s)\bp
J_{\pi\eta}(s) & 0 \\[0.2cm]
0 & J_{KK}(s)      \\ \ep
\en
where $\bm{T}^{(2)}(s)$ is the chiral $O(p^2)$ $T$-matrix and
$J_{P_1P_2}(s)$ are one-loop functions which satisfy
\be
\im[J_{P_1P_2}(s)]= \sigma_{P_1P_2}(s)\ .
\en
In the Omn\`es representations we had separated the Born amplitude
into a singular and a regular part
\be
k^{1,B}_{0++}(s)=\alpha_B\left( \frac{\beta}{s-\qd}+\gamma\right)
+\bar{k}^{1,B}_{0++}(s),
\en
we can first express the right-cut integrals as
\be
I^{RC}_i(s,\qd)=\frac{s-s_0}{\pi}\int_{4\mkd}^\infty ds'
\frac{ (k^{1,B}_{0++}(s')-\bar{k}^{1,B}_{0++}(s')) T^{(2)}_{i2}(s')\,\sigma_K(s')}
{(s'-s_0)(s'-\qd)(s'-s)}
\en
using the unitarity relation for the matrix elements
$D_{i2}(s)$ 
\be\lbl{imDi2simple}
\im[D_{i2}(s')] = -(\bm{D}(s')\bm{T}(s'))_{i2}\,\sigma_K(s')
= -T^{(2)}_{i2}(s')\,\sigma_K(s')\ 
\en
where the first equality is general and the second one follows
from~\rf{Dmatrixoset}.  The fact that $T^{(2)}_{i2}(s')$ are linear functions
of $s'$ is the key to the simplification.  Indeed, we can express the Born
left-cut integrals using a dispersive representation for the product
$\bar{k}^{1,B}_{0++}(s)D_{i2}(s)/(s-\qd)$ (which remains finite when $s=\qd$)
obtaining
\be
I_i^{LC,Born}(s,\qd)=\frac{\bar{k}^{1,B}_{0++}(s)D_{i2}(s)}{s-\qd}
-\frac{s-s_0}{\pi}\int_{4\mkd}^\infty ds' 
\frac{ \bar{k}^{1,B}_{0++}(s')\im[D_{i2}(s')]}{(s'-s_0)(s'-\qd)(s'-s)}
\en
Then, replacing $\im[D_{i2}(s')]$ with~\rf{imDi2simple} and adding the
two integrals one gets
\be
\ba{l}
(s-\qd)\left(I_i^{LC,Born}(s,\qd)+ I_i^{RC}(s,\qd) \right)= \\[0.3cm]
\qquad \bar{k}^{1,B}_{0++}(s)D_{i2}(s)
+\dfrac{(s-\qd)(s-s_0)}{\pi}{\displaystyle\int_{4\mkd}^\infty}ds' 
\dfrac{k^{1,B}_{0++}(s') T^{(2)}_{i2}(s')\,\sigma_K(s')}
{(s'-s_0)(s'-\qd)(s'-s)}\ .
\ea\en
We can then choose the subtraction point $s_0$ such that 
$T^{(2)}_{i2}(s')/(s'-s_0)$ is a constant, which effectively pulls out
$T^{(2)}_{i2}(s') $ from the integral. The remaining integral is
equal to the kaon loop function $I^{loop}_K(s,\qd)$. Finally, one arrives at
the following form for the two $\gamma\gamma^*$ amplitudes,
\be
\bp l_{0++}(s)\\[0.3cm]
k^1_{0++}(s)\\ \ep=
\bp 0 \\[0.3cm]
k^{1,B}_{0++}(s)\ep
+ (s-\qd)\bm{\Omega}(s) \bp a_1 \\[0.3cm]
b_1\ep
+ I^{loop}_K(s,\qd)
\bp T_{12}(s) \\[0.3cm]
T_{22}(s) \\ \ep
\en
using that $\bm{\Omega}(s)\bm{T}^{(2)}(s)= \bm{T}(s)$ in this model.
The integrals associated with meson exchanges e.g. $I^{LC,V}(s)$ in
the Omn\`es representation can be similarly re-expressed in terms of
simple triangle loop functions using eq.~\rf{imDi2simple} which allows
one to recover the formulae of ref.~\cite{Palomar:2003rb}.

\section{$\rho$-meson spectral function}\lblsec{roV}
Integrals involving the spectral function $\rho^V$ which describe the
width of the $\rho$ meson, in particular $\rho^V_1$, $\rho^V_2$
(eqs.~\rf{rhoV1rhoV2}), are easily computed in the model given by
eq.~\rf{rhoVmodel} by writing the denominator of $\rho^V$, which is a
cubic polynomial,  as a product of three factors,
\be
D_V(t)=(1+\gamma_V^2)(t-z_R)(t-z_+)(t-z_-)
\en
where $z_R$ is a real root and $z_\pm$ are complex conjugated. These
roots must be computed numerically. The integrals of interest involve
the following functions
\be
\ba{l}
F_R(t)=(t_0-z_R)^{3/2}\arctan\dfrac{\sqrt{t-t_0}}
{\sqrt{t_0-z_R}},\\[0.4cm]
F_\pm(t)=\frac{1}{2}(z_\pm-t_0) ^{3/2}
      \log\dfrac{-\sqrt{t-t_0} + \sqrt{z_\pm-t_0}}
                {\sqrt{t-t_0} + \sqrt{z_\pm-t_0}}\ .
\ea\en
For instance, one has
\be\lbl{rho0V}
\int_{t_0}^t   dt' \rho_V(t';m_V,\Gamma_V)=
\frac{2N_V\gamma_V}{1+\gamma_V^2}
\left( \alpha F_R(t) +\beta F_+(t) +\beta^* F_-(t)
\right)
\en
with
\be
\alpha=\frac{1}{|z_+ -z_R|^2},\quad
\beta=\frac{1}{(z_+-z_R)(z_+-z_-)}\ .
\en
Taking the limit $t\to\infty$, one can determine $N_V$ from the normalisation
condition~\rf{normalroV} which yields 
\be
N_V=\frac{1+\gamma_V^2}{\pi\gamma_V}
\left(\frac{(t_0-z_R)^{3/2}}{|z_+-z_R|^2} 
-\re\left[ \frac{(z_+-t_0)^{3/2}}{(z_+-z_R)\im[z_+]}\right] \right)^{-1} .
\en
We can express in a similar way the following two integrals,
\be\lbl{rhobarVdef}
\bar{\rho}^1_V\equiv \int_{t_0}^t dt' t' \rho^V(t';m_V,\Gamma_V),\quad
\bar{\rho}^2_V\equiv \int_{t_0}^t dt'(t'-\metad)^2 \rho^V(t';m_V,\Gamma_V),\quad
\en
as
\be\lbl{rho1V}
\bar{\rho}^V_1(t)=\frac{2N_V\gamma_V}{1+\gamma_V^2}\left( 
\sqrt{t-t_0}+ \alpha' F_R(t) +\beta' F_+(t) +{\beta'}^* F_-(t)\right)
\en
with
\be
\alpha'=\alpha z_R,\quad \beta'=\beta z_+ \ ,
\en
and
\be\lbl{rho2V}
\ba{ll}
\bar{\rho}^V_2(t)=& \dfrac{2N_V\gamma_V}{1+\gamma_V^2}\Big( 
\sqrt{t-t_0}\left(\frac{1}{3}(t-4t_0) 
+\alpha''z_R +\beta''z_++{\beta''}^*z_- \right)\\[0.4cm]
\ & + \alpha'' F_R(t) +\beta'' F_+(t) +{\beta''}^* F_-(t) \Big)
\ea\en
with
\be
\alpha''= \alpha(z_R-\metad)^2,\quad
\beta''=\beta(z_+-\metad)^2\ .
\en
The integrals $\rho^V_1$, $\rho^V_2$ introduced in eq.~\rf{rhoV1rhoV2}
are simply related to $\bar{\rho}^V_1$, $\bar{\rho}^V_2$,
\be
\rho^V_1(t)=\bar{\rho}^V_1(t)-\bar{\rho}^V_1(\metad),\quad
\rho^V_2(t)=\bar{\rho}^V_2(t)-\bar{\rho}^V_2(\metad)\ .
\en
Finally, using this model, the  propagator of the
$\rho$ meson (eq.~\rf{kallenlehman}) is expressed as a sum of four terms
\be
\ba{ll}
P_\rho(u;m_V,\Gamma_V)=&
N_V\dfrac{\pi\gamma_V}{1+\gamma_V^2}\Big\{ 
\dfrac{ -(t_0-u)^{3/2}}{(u-z_R)(u-z_+)(u-z_-)}\\[0.4cm]
\ & +\alpha\dfrac{(t_0-z_R)^{3/2}}{u-z_R}
+\beta\dfrac{ (t_0-z_+)^{3/2}}{u-z_+}
+\beta^*\dfrac{ (t_0-z_-)^{3/2}}{u-z_-}
\Big\}\ .
\ea
\en

\section{$K$-matrix parametrisation}\lblsec{K-matrix}
We briefly recall below the parametrisation of the two-channel
($\pi\eta$, $K\Kbar$) $S$-wave scattering amplitudes introduced in
ref.~\cite{Albaladejo:2015aca}. The $T$ matrix is expressed in terms
of a $2\times2$ symmetric $K$-matrix,
\be
\bm{T}(s)=(\bm{1}-\bm{K}(s)\bm{\Phi}(s))^{-1}\bm{K}(s)
\en
Two-channel unitarity is satisfied, as is well known, provided that
the $K$-matrix elements are real in the physical region $s \ge
(\meta+\mpi)^2$ and the $\Phi$-matrix satisfies
\be\lbl{imPhi}
\im[\bm{\Phi}(s)]=\bp
\theta(s-(\meta+\mpi)^2)\sigma_{\pi\eta}(s) & 0 \\
0 & \theta(s-4\mkd) \sigma_K(s) \\
\ep
\en
We use a $\Phi$-matrix which involves four phenomenological parameters
\be
\bm{\Phi}(s)=\bp
\alpha_1 +\beta_1 s+ 16\pi \bar{J}_{\pi\eta}(s) & 0 \\
0 & \alpha_2 +\beta_2 s+ 16\pi \bar{J}_{K\Kbar}(s) \\ 
\ep
\en
where $\bar{J}_{P_1P_2}(s)$ are one-loop functions as defined in
ref.~\cite{Gasser:1984gg}, which satisfy $\bar{J}_{P_1P_2}(0)=0$ and
$\im[16\pi \bar{J}_{P_1P_2}(s)]=\sigma_{P_1P_2}(s)$. The parameters
$\alpha_i$, $\beta_i$ are real and assumed to be of chiral order
$p^0$.

The $K$-matrix is written as a sum of three terms of increasing chiral
order $\bm{K}(s)= \bm{K}^{(2)}(s)+\bm{K}^{(4)}(s)+\bm{K}^{(6)}(s)$ in
which the terms of order $p^2$ and $p^4$ are determined by matching to
the chiral expansion of the $T$-matrix i.e.
\be
\bm{K}^{(2)}=\bm{T}^{(2)},\quad
\bm{K}^{(4)}=\bm{T}^{(4)}- \bm{T}^{(2)}\bm{\Phi}^{(0)}\bm{T}^{(2)}\ 
\en
where $\bm{\Phi}^{(0)}$ is the $O(p^0)$ part of $\bm{\Phi}$. 
The chiral expansions of $T_{ij}$ up to NLO were first derived
in~\cite{Bernard:1991xb} and in~\cite{Guerrero:1998ei,GomezNicola:2001as}.  The
matching can be performed exactly for the matrix elements $T_{11}$,
$T_{12}$ but only approximately for $T_{22}$ in order to have $K_{22}$
real. The matrix elements ${K}^{(4)}_{ij}$ depend on the
Gasser-Leutwyler~\cite{Gasser:1984gg} coupling constants $L^r_i$. In
this work, their values were taken from table 3 of
ref.~\cite{Bijnens:2014lea} (BE14 set).
The third term,
$\bm{K}^{(6)}$, implements a $K$-matrix pole 
\be\lbl{K6}
\left(\bm{K}^{(6)}(s)\right)_{ij}=\frac{\lambda g_i g_j}{16\pi}
\left( \frac{1}{m_8^2-s}-\frac{1}{m_8^2}
\right)
\en
where $\lambda$ and $m_8$ are phenomenological parameters while the
form of $g_i$ is derived from a
resonance chiral Lagrangian\cite{Ecker:1988te}
\be\lbl{g1g2}
\ba{l}
g_1=\dfrac{1}{\sqrt3\fpid}(c_d (s-\Sigma_{\pi\eta})+2 c_m \mpid)\\[0.3cm]
g_2=\dfrac{1}{\fpid}(c_d(s-2\mkd)+ 2c_m \mkd)\ .
\ea
\en
The central values of $c_m$, $c_d$ were taken as
\be
c_d=28\,\, \hbox{MeV},\quad
c_m=2c_d\ .
\en
The form described above is used in a limited energy region 
$s \le s_{cut}$ with $\sqrt{s_{cut}}=1.5$ GeV. The $T$-matrix must also
be defined for  $s > s_{cut}$  in order to solve for the 
$\bm{\Omega}$ matrix. In this region, a simple
interpolation of the two phase shifts (such that the sum goes to
$2\pi$ at infinity) and of the inelasticity (going to 1 at infinity)
is implemented.  
We collect the numerical values of the sets of parameters
corresponding to the fits described in sec.~\sect{a0results} in
table~\Table{paramvalues} below. 
\begin{table}
\centering \bt{l| c c c}\hline\hline 
\  & $\gamma\gamma$ fit &  $\gamma\gamma$ fit   & $\gamma\gamma+ \phi$ fit\\
\  & $(s^{\gamma\gamma}_A=\metad)$     &
$(s^{\gamma\gamma}_A=\metad+3\mpid)$ & 
$(s^{\gamma\gamma}_A=\metad+3\mpid)$   \\ \hline 
$\alpha_1$ & 0.996820807   &    0.984091282     &     1.0386891\\
$\alpha_2$ &-0.494739205   &   -0.495522797    &    -0.48956928\\
$\beta_1$ (GeV$^{-1}$)&-4.14119816  &    -3.78536367  &  -4.5487204\\
$\beta_2$ (GeV$^{-1}$)&-0.185484648 &    -0.120648287 &  -0.16840835\\
$m_8$  (GeV)          &0.899005413 &     0.910190940  &   0.88816762\\
$\lambda$ &1.06326795    &   1.07898855	  &     1.0276550\\ \hline\hline 
\et 
\caption{\small  Numerical  values of the  phenomenological $T$-matrix
parameters corresponding to various fits discussed in sec.~\sect{datafits}.}
\lbltab{paramvalues} 
\end{table}

\section{Estimate of the $a_2(1320)$ contribution}\lblsec{a2amplit}
The contributions to the $\phi\to \gamma\pi\eta$ helicity amplitudes
induced by the $a_2(1320)$ tensor resonance, i.e. $\phi\to
\gamma a_2(1320) \to \gamma\pi\eta$ have the following form,
\be\lbl{tensora2amplit}
\ba{l}
L^T_{++}(s,\theta)= C^T \dfrac{\qd (\qd-s)\lambda^2_{\pi\eta}(s)}
{48 s^2 (m_T^2 -s -im_T\Gamma_T)}(3\cos^2\theta-1)\\[0.3cm]
L^T_{+-}(s,\theta)= C^T \dfrac{(\qd-s) \lambda^2_{\pi\eta}(s)}
{16\,s (m_T^2 -s -im_T\Gamma_T)}\sin^2\theta\\[0.3cm]
L^T_{+0}(s,\theta)=  C^T\dfrac{\sqrt{2 \qd} (\qd-s) \lambda^2_{\pi\eta}(s)}
{16\sqrt{s}\, s (m_T^2-s -im_T\Gamma_T)} 
\sin\theta\cos\theta
\ea\en 
where the constant $C^T$ is proportional to the product of the $a_2\pi\eta$
and $a_2\phi\gamma$ couplings. Unfortunately, there is no experimental
information on $a_2\to \phi\gamma$ decay. It is possible to get a qualitative
estimate of the $a_2\phi\gamma$ coupling, relating it to the known
$a_2\to 2\gamma$ coupling, using vector meson dominance ideas,
following ref.~\cite{Renner:1971sj}. This gives
\be
C^T\simeq -\frac{3e\,C^{a_2}_{\gamma\gamma}\,C^{a_2}_{\pi\eta}}
{2f_V}\,s_1
\en
where the coupling constants  $C^{a_2}_{\gamma\gamma}$, $C^{a_2}_{\pi\eta}$, 
defined in~\cite{Lu:2020qeo}, have the values
\be
\ba{ll}
C^{a_2}_{\gamma\gamma}=& (0.115\pm 0.005)\ \hbox{GeV}^{-1}\\[0.2cm]
C^{a_2}_{\pi\eta}   =& (10.8\pm0.5) \hbox{GeV}^{-1}\ .
\ea\en
According to this estimate, one finds that the contribution from the $a_2$
resonance is rather small as compared to the vector-exchange
contributions. For instance, at $s=0.98$ $\hbox{GeV}^2$, defining the ratios 
$R_{\lambda\lambda'}= \sum_V L^V_{\lambda\lambda'}/ L^T_{\lambda\lambda'}$ one
obtains
\be
R_{++}=(1.0-i0.4)\cdot10^{-2},\
R_{+-}=(6.5-i0.7)\cdot10^{-2},\
R_{+0}=(3.3-i1.0)\cdot10^{-2}\ .
\en


\end{document}